\documentclass[twocolumn,aps,pra,showpacs,
fixfloats]{revtex4-1}
\usepackage{bm}
\usepackage{dcolumn}
\usepackage{graphicx}
\begin{document}
\newcommand{\ra}{\rangle}
\newcommand{\la}{\langle}
\renewcommand{\k}{{\bbox k}}
\newcommand{\hs}{\hspace*}
\newcommand{\vs}{\vspace*}
\newcommand{\np}{\newpage}
\newcommand{\cx}{C$_{60}$~}
\newcommand{\eref}[1] {(\ref{#1})}
\newcommand{\Eref}[1] {Eq.~(\ref{#1})}
\newcommand{\Fref}[1] {Fig. \ref{#1}}
\newcommand{\nn}{\nonumber}
\newcommand{\be}{\begin{equation}}
\newcommand{\ee}{\end{equation}}
\newcommand{\br}{\begin{eqnarray*}}
\newcommand{\er}{\end{eqnarray*}}
\newcommand{\ba}{\begin{eqnarray}}
\newcommand{\ea}{\end{eqnarray}}
\newcommand{\bp}{\begin{minipage}}
\newcommand{\ep}{\end{minipage}}
\newcommand{\ds}{\displaystyle}
\newcommand{\bs}{\bigskip}

\title{Polarization of the C$_{60}$ cage by the photoelectron, interior static polarization of C$_{60}$ by the ion-remainder and atomic-core relaxation in $A$@C$_{60}$ photoionization}
\author{ V. K. Dolmatov}
\email[Send e-mail to:\ ]{vkdolmatov@una.edu}
\address{Department of Physics and Earth Science, University of North Alabama,
Florence, Alabama 35632, USA}
\begin{abstract}
Photoionization cross sections, $\sigma_{n\ell}$, photoelectron angular-asymmetry parameters, $\beta_{n\ell}$, and photoionization time delays, $\tau_{n\ell}$, for endohedral atoms, $A$@C$_{60}$, are studied with account for both the individual and combined
effects of dipole static polarization (DSP) of the C$_{60}$ fullerene cage by the outgoing photoelectron, interior static polarization (ISP) of C$_{60}$ by the ion-remainder, $A^{+}$,
and atomic-core relaxation of the encapsulated atom upon its ionization. It is unraveled by the direct calculations that the DSP effect is weak. It changes the phase of confinement-resonant
oscillations in $\sigma_{n\ell}$, $\beta_{n\ell}$ and $\tau_{n\ell}$ without generally noticeable changes in their magnitudes, unless $\sigma_{n\ell}$ concentrates a relatively large part of oscillator strength
of the continuum spectrum near threshold; then and there, the changes can be noticeable. This is counter-intuitive  in view of a large dipole static polarizability of C$_{60}$, $\alpha > 800$ $a.u.$. Furthermore,
it is demonstrated that the DSP effect results in the transmission of a part of oscillator strength of the continuum spectrum of $A$@C$_{60}$ into its discrete spectrum. It is shown that the DSP effect is counteracted by ISP, thus getting partially cancelled out by the latter. Possible reasons behind the made findings are provided. Photoionization of Xe@C$_{60}$, Ne@C$_{60}$, H@C$_{60}$ and some hypothetical C$^{-}_{60}$ fullerene anions is chosen as a case study. For Xe@C$_{60}$, the role of atomic-core relaxation of the ionized encapsuled Xe$^{+}$-ion-remainder on the $\rm 4d$ photoionization of Xe@C$_{60}$ is explored and detailed, and is revealed to be of utter significance, as in the known case of free Xe. The study is performed in the frameworks of both the random phase approximation with exchange (RPAE) and
generalized RPAE (GRPAE), when needed. The C$_{60}$ cage is modelled by a spherical attractive potential of a certain inner radius, thickness and depth. Its dipole static polarization potential is approximated by the Bates dipole static potential.
\end{abstract}

\maketitle
\section{Introduction}

Photoionization of atoms, $A$, encapsulated inside a C$_{60}$ fullerene -- the $A$@C$_{60}$ endohedral atoms -- has been a topic of many studies by theory (see, e.g., Refs.~\cite{PushkaPRA93,Balt1,Balt2,JPC,Lo,Korol,Amusia@Xe,Amusia@GR,JPCVKD,DolmCCR,Dolm@RR,DolmAQC,DolmPRA2010,Ludlow,DolmW-S,Lee,DeshmukhJPB09,Deshmukh@TD,Himadri1,Himadri2,GrumJPB11,
Jose,Drukarev,Fricke,Keating,Martinez,Phaneuf1}, and references therein) and, since recently, experiment \cite{Phaneuf1,Phaneuf2}.
One of the fascinating features of $A$@C$_{60}$ photoionization  is the existence of resonances termed the confinement resonances, that occur due to the
interference of the photoelectron waves emerging directly from
the encapsulated (confined) atom and those scattered off the C$_{60}$ confining cage. Their existence has recently been confirmed experimentally  \cite{Phaneuf1,Phaneuf2}.

Many of theoretical results on this topic have been obtained in the framework of a model potential where the C$_{60}$ carbon cage is modelled by a
Dirac $\delta$-like potential \cite{Balt1,Amusia@Xe,Amusia@GR} (and references therein), or a $U_{\rm C}(r)$ spherical attractive potential of a certain inner radius, thickness and depth
\cite{JPCVKD,DolmCCR,Dolm@RR,DolmAQC,DolmPRA2010,Ludlow,DeshmukhJPB09,Deshmukh@TD,GrumJPB11,Fricke,Keating,Martinez,Phaneuf1}, or Woods-Saxon potential \cite{DolmW-S,Lee,Martinez}, or other potentials \cite{Balt2}.
However, all of these studies have overlooked to account for polarization of C$_{60}$ by the outgoing photoelectron (to be referred to as ``$\alpha$-polarization'', $\alpha$ being dipole static polarizabity). Meanwhile, it has recently been found \cite{Dolm@Ba+e-} that $\alpha$-polarization of  C$_{60}$ has a drastic impact on electron scattering both off empty
 C$_{60}$ and $A$@C$_{60}$, resulting not only in quantitative changes of corresponding scattering phases and cross sections, but in their qualitative changes as well, e.g., bringing Ramsauera-type
 minima in the scattering cross sections. Now,
 upon photoionization of $A$@C$_{60}$, the outgoing photoelectron, too, scatters off the C$_{60}$ cage on its way out of
  $A$@C$_{60}$. Hence, photoionization of $A$@C$_{60}$  depends on polarizability of C$_{60}$ as well, somehow. But... how?  Our study provides the insight into this problem. We choose the $\rm 4d$ photoionization of Xe@C$_{60}$,
  $\rm 2p$ photoionization of Ne@C$_{60}$, $\rm 1s$ photoionization of H@C$_{60}$ and photodetachment of hypothetical C$_{60}^{-}(\rm 2p)$ and C$_{60}^{-}(\rm 3d)$  fullerene anions as case studies, and calculate corresponding photoionization cross sections, $\sigma_{n\ell}$, dipole angular asymmetry parameters, $\beta_{n\ell}$, and photoionization time delay \cite{Schultze}, $\tau_{n\ell}$, both with and without account for the  $\alpha$-polarization impact on the photoionization process.

  Furthermore,  we also explore and detail the impacts of other effects on the photoionization process. One of them is termed ``interior static polarization'' of C$_{60}$ \cite{DolmPRA2010}. This is the effect of static polarization of the C$_{60}$ fullerene cage caused not by an outgoing photoelectron but by the $A^{+}$ ion-remainder in response to photoionization of $A$@C$_{60}$. This effect depends on some parameter $\zeta$, thus suggesting the name ``the $\zeta$-polarization effect'' which (the name) will be used throughout the paper. The other effect is the correlation-related impact of atomic-core relaxation of the encapsulated ion-remainder, $A^{+}$, upon $A$@C$_{60}$ photoionization. Atomic-core relaxation is known to be decisive for the formation of $\rm 4d$ photoionization of free Xe \cite{Amusia}. However, how this correlation-related relaxation effect will be developing in Xe@C$_{60}$ is not a priory certain, since correlation may work quite differently in endohedral atoms compared to free atoms \cite{JPCVKD}.

  To account for electron correlation in photoionization of Xe@C$_{60}$, we utilize both the random phase approximation with exchange (RPAE) and generalized RPAE (GRPAE) \cite{Amusia}. GRPAE not only accounts for electron correlation, as RPAE, but, as well, includes the effect of atomic-core relaxation as the photoelectron leaves the atom. The overall agrement between the calculated GRPAE  $\sigma_{4d}$ of Xe@C$_{60}$ and experiment \cite{Phaneuf1} is found to be good, in contrast to calculated RPAE $\sigma_{4d}$. This reveals the importance of the effect of atomic-core relaxation not only in free Xe but in Xe@C$_{60}$ photoionization as well -- a \textit{particular} effect which remains to be due to the specificity of Xe.

In order to elucidate the impacts of the presence of the C$_{60}$-cage environment and its polarization by the photoelectron and/or by the $A^{+}$ ion-remainder on the $A$@C$_{60}$ photoionization process, we approximate the resultant C$_{60}$  potential, ``felt'' by the outgoing photoelectron, as the sum of the $U_{\rm C}(r)$ spherical potential, the Bates static polarization potential, $V_{\alpha}$ \cite{Bates}, that depends on the dipole static polarizability, $\alpha$, of C$_{60}$ and
the $V_{\zeta}$ potential caused by interior static polarization \cite{DolmPRA2010}. These potentials are added to the Hartree-Fock (HF) atomic potential of the encapsulated atom in order to account for the C$_{60}$ polarizable environment.

Speaking about Xe@C$_{60}$, a surprising
result of the study is that the impact of  polarization of C$_{60}$ by the outgoing photoelectron (the $\alpha$-polarization impact)  on $\sigma_{\rm 4d}$, $\beta_{\rm 4d}$ and $\tau_{\rm 4d}$ of Xe@C$_{60}$
 is found to be relatively weak despite of a large value of C$_{60}$'s $\alpha = 850$ $a.u.$ \cite{Amusia_alpha}. Corresponding calculated data for Xe@C$_{60}$, obtained both   with  and without account for $V_{\alpha}$ (i.e., with and without account for $\alpha$-polarization), do not differ significantly from each other; the corresponding confinement resonant oscillations are shifted  by about $2$ eV towards threshold (i.e., their phase is changed) when $\alpha$-polarization of C$_{60}$ is taken into account in the calculation. Without strong exaggeration, one may say that the graphs for  $\sigma_{\rm 4d}$, $\beta_{\rm 4d}$ and $\tau_{\rm 4d}$ appear to be shifted, practically as a whole, towards threshold, and their parts, that go beyond threshold, are simply cut off the rest part of each graph. It is in this sense that we will refer to this situation as the shift of $\sigma_{n\ell}$, $\beta_{n\ell}$ and $\tau_{n\ell}$ towards threshold.

 Speaking about Ne@C$_{60}$ and H@C$_{60}$, the same energy shift, due to $\alpha$-polarization, is found to take place for their photoionization spectra as well. However, in contrast to Xe@C$_{60}$,
 the $\alpha$-polarization effect causes noticeable changes in $\beta_{\rm 2p}$ of Ne@C$_{60}$, near threshold, and in $\sigma_{\rm 1s}$ of H@C$_{60}$, near threshold; reasons will be provided later in the paper. Here, we only point out that, based on those reasons, it is our current understanding that the $\alpha$-polarization effect is important if the photoionization cross section contains a relatively large part of oscillator strength of the continuum spectrum near threshold, within a narrow photon energy region of about $2$ eV starting from threshold; then and there, the $\alpha$-polarization impact on near-threshold photoionization should be noticeable.

  Furthermore, it is uncovered that the effect of interior static polarization of C$_{60}$ (the $\zeta$-polarization effect) noticeably counter-acts the $\alpha$-polarization effect. As a result, when both effects are
  accounted in the calculation of $A$@C$_{60}$ photoionization, the resultant spectrum differs only relatively insignificantly from the spectrum calculated without account for both polarization effects altogether, i.e., the entire polarization effect can be ignored, to a good approximation, unless one focuses on near-threshold $\sigma_{n\ell}$, $\beta_{n\ell}$ and $\tau_{n\ell}$.

  Atomic units ($a.u.$) are used throughout the paper unless stated otherwise.

\section{Review of Theory, Results and Discussion}

\subsection{Basic formulas}

To calculate photoionization quantities of $A$@C$_{60}$, we utilize the RPAE and GRPAE theories \cite{Amusia}.

Let us first comment on RPAE and GRPAE with regard to a free atom.

\begin{figure}
\includegraphics[width=7cm]{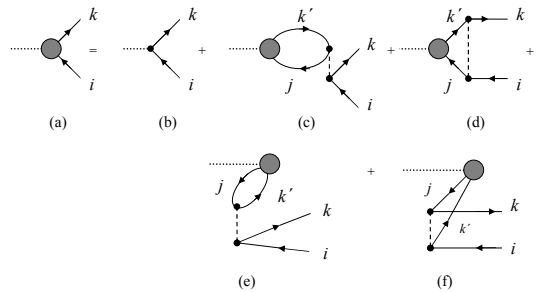}
\caption{Feynman-Goldstone
diagrammatic representation of the RPAE equation for the
photoionization amplitude $\la k|\hat{D}|i\ra $ of the \textit{i}'th subshell
into the \textit{k}'th final state. Here, the time axis is directed from the left to right,
the lines with
arrows to the left (right) correspond to holes (electrons) in the atom, a
dotted line represents an incoming photon, a dashed line represents the
Coulomb interaction $V(r)$ between charged particles, and a shaded circle marks
 the effective operator $\hat{D}$ for the photon-atom interaction which accounts for electron
correlation in the atom.}
\label{fig1}
\end{figure}

RPAE \cite{Amusia} uses the Hartree-Fock  (HF) basis as the zeroth-order basis (the vacuum state) and accounts for electron correlation in photoionization amplitudes, $D_{n\ell \rightarrow \ell\pm 1}$, by summing up a certain infinite series of Feynman-Goldstone many-body diagrams, depicted graphically in Fig.~\ref{fig1}.
There, diagrams (c)--(f) represent RPAE corrections to the
HF photoionization amplitude $\la k|\hat{d}|i\ra \equiv d_{ik}$
[diagram (b)]. The analytical expression for the RPAE equation is cumbersome; we refer the reader to Ref.~\cite{Amusia} for details.

The RPAE photoionization cross section, $\sigma_{n\ell}$, of a $n\ell^{2\ell+1}$ subshell of the atom and corresponding $\beta_{n\ell}$ angular-asymmetry parameter are determined by known formulas,
see, e.g., Ref.~\cite{Amusia}:
\begin{eqnarray}
 \sigma_{n \ell} = \frac{4\pi{^2}\alpha N_{n \ell}}{3(2\ell+1)} \omega \left[\ell
|D_{\ell-1}|^{2}+ (\ell+1)|D_{\ell+1}|^{2}\right],
\label{eqCC}
\end{eqnarray}
and we recast $\beta_{n\ell}(\omega)$ as
\begin{eqnarray}
 \beta_{n \ell} =
\frac
{
\ell (\ell-1)\rho^{2} - 6\ell(\ell+1)\rho\cos\Phi
+ (\ell +1)(\ell+2)}
{(2\ell+1)(\rho^{2}\ell +\ell +1)
},
\label{eqbeta2}
\end{eqnarray}
where
\begin{eqnarray}
\rho = \frac{|D_{\ell-1}|}{|D_{\ell+1}|}, \quad \Phi=\delta_{\ell+1}-\delta_{\ell-1}.
\label{eqrho}
\end{eqnarray}
In the above equations, $\omega$ is the photon energy, $N_{n\ell}$ is the
number of electrons in a $n\ell$ subshell, $\alpha$ is the fine-structure constant,
$D_{\ell\pm 1}$ are radial parts of reduced dipole photoionization amplitudes (which are complex in the RPAE theory) and $\delta_{\ell \pm 1}$ are phases of the $D_{\ell\pm 1}$ amplitudes.
 Note that the quantities $\sigma_{n\ell}$, $\beta_{n\ell}$, $D_{\ell\pm 1}$,
  $\rho$, $\delta_{l\pm1}$, and
$\Phi$ all depend on photon energy $\omega$; the explicit dependence
is omitted in the above equations for reasons of simplicity.

RPAE utilizes the excited atomic states calculated in the field
of the frozen atomic core with a whole in the ionized subshell $n\ell$. RPAE, thus, neglects the effects of rearrangement (relaxation) of the atomic core while the outgoing photoelectron leaves the atom. In some cases, like the
$\rm 4d$ photoionization of Xe, the effect of relaxation of the atomic-core on $n\ell$ photoionization is known to be crucial \cite{Amusia}. This effect is well accounted for by generalized RPAE, i.e., GRPAE.
GRPAE differs from RPAE in that the wavefunction of an outgoing photoelectron, to be used in GRPAE, is calculated not in the field of a frozen ion-remainder $A^{+}$, but in the self-consistent field of the completely rearranged (relaxed) ion. In the case of the $\rm 4d$ photoionization of Xe, the impact of the relaxed Xe$^{+}(...4d^{9}...)$ ion on the photoionization process is of utter importance and results in a good agreement between calculated GRPAE and experimental $\rm 4d$-photoionization cross section and $\beta_{4d}$ angular-asymmetry parameter for free Xe \cite{Amusia}. However, as was pointed above, how this correlation-related relaxation effect will develop in Xe@C$_{60}$ is not a priory certain, since correlation may work quite differently in endohedral atoms compared to free atoms \cite{JPCVKD}.

We now briefly outline the key points of the Eisenburg-Wigner photoionization time
delay concept which is the subject of our study as well. Similar to the Eisenburg-Wigner theory for time
delay in electron scattering \cite{Wigner}, time
delay in the photoionization of a $n_{i}\ell_{i}$ subshell of the atom is known to be
determined by a derivative of the phase $\varphi(E)$ of corresponding
photoionization amplitude $D_{n_{i}\ell_{i}}=
\left|D_{n_{i}\ell_{i}}\right|e^{i\varphi(E)}$:
\begin{eqnarray}
 \varphi(E) = \arg[D_{n_{i}\ell_{i}}(E)], \quad \tau_{n_{i}\ell_{i}}= \frac{d\varphi(E)}{dE}.
\label{phase}
\end{eqnarray}
For a photoionization amplitude, $D_{n_{i}\ell_{i}}$, of a $n_{i}\ell_{i}$-state, which accounts for both
$n_{i}\ell_{i} \to \epsilon(\ell_{i} \pm 1)$ dipole transitions, one has
\cite{Kheifets}:
\begin{eqnarray}
\label{amplitude}
D_{n_i \ell_i}(E)&\propto&
\sum_{\ell=\ell_i\pm1\atop m=m_i}
e^{i\delta_l}i^{-l}
Y_{lm}(\hat{\bm k})\,
(-1)^m
\left(\begin{array}{rrr}
\ell&1&\ell_i\\
-m&0&m_i\\
\end{array}\right)
\nn\\&&\hs{2cm}\times \
 \la E\ell\|D\|n_i \ell_i \ra.
 \label{Dnl}
\end{eqnarray}
Here, $\hat {\bm k}$ is a unit vector in the direction of the
photoelectron momentum $\bm k$, $\delta_{\ell}(E)$ is the phase shift of
the $\ell$th outgoing photoelectron wave, and $\la E\ell\|D\|n_i \ell_i \ra$ is
the reduced dipole matrix element
which is the solution of the RPAE equation, Fig.~\ref{fig1}.
In the present work, $D_{n_{i} \ell_{i}}(E)$ is evaluated in the forward direction
${\bm k}\|\hat z$.

To apply the above formulas to photoionization of $A$@C$_{60}$, we need to determine the ground-state and excited states of the atom encapsulated inside C$_{60}$. We do it following the strategy developed in
numerous previous works, see, e.g., Refs.~\cite{JPCVKD,DolmCCR,Dolm@RR,DolmAQC,DolmPRA2010,Ludlow,DeshmukhJPB09,Deshmukh@TD,GrumJPB11,Fricke,Keating,Martinez,Phaneuf1}.

First, we model the C$_{60}$ cage by a $U_{\rm C}(r)$ spherical potential:
\begin{eqnarray}
U_{\rm C}(r)=\left\{\matrix {
-U_{0}, & \mbox{if $r_{0} \le r \le r_{0}+\Delta$} \nonumber \\
0 & \mbox{otherwise.} } \right.
\label{SWP}
\end{eqnarray}
Here, $r_{0}$, $\Delta$, and $U_{0}$ are the inner radius, thickness, and depth of the potential well, respectively.

Second, if the encapsulated atom, $A$, is compact, the atom $A$ and C$_{60}$ can be regarded as independent entities. Then, to calculate the ground-state of such endohedral atom, we simply add the $U_{\rm C}$ cage potential
to the Hartree-Fock (HF) of the encapsulated atom, thereby producing the ``endohedral'' HF equations:
\begin{eqnarray}
&&\left[ -\frac{\Delta}{2} - \frac{Z}{r} +U_{\rm c}(r) \right]\psi_{i}
({\bm x}) + \sum_{j=1}^{Z} \int{\frac{\psi^{*}_{j}({\bm x'})}{|{\bm
x}-{\bm x'}|}} \nonumber \\
 && \times[\psi_{j}({\bm x'})\psi_{i}({\bm x})
- \psi_{i}({\bm x'})\psi_{j}({\bm x})]d {\bm x'} =
\epsilon_{i}\psi_{i}({\bm x}).
\label{eqHF}
\end{eqnarray}
Here,  $\psi_{n \ell m_{\ell} m_{s}}({\bm r}, \sigma)=r^{-1}P_{nl}(r)Y_{l m_{\ell}}(\theta, \phi) \chi_{m_{s}}(\sigma)$ are the electronic wavefunctions,
$\epsilon_{n l}$ are electronic energies,
($n$, $\ell$,  $m_{\ell}$ and $m_{s}$ is the standard set of quantum numbers of an electron in a central field, $\sigma$ is the electron spin coordinate). Once the ground-state wavefunctions of $A$ are determined, let us
assign them a subscript $j$, they are plugged back into the equation to determine the excited-state wavefunctions, $\psi_{i}({\bm x})$ (we assign a subscript $i$ to them). This procedure, employed
in all previous studies cited above, actually, defines the wavefunction of an outgoing photoelectron without account for the effect of polarization of C$_{60}$ by the photoelectron itself.

To account for the $\alpha$-polarization effect in question, let us build a $U_{\rm C\alpha}(r)$ potential of a \textit{polarized} C$_{60}$, ``felt'' by a photoelectron, as the sum of the $U_{\rm C}(r)$ cage potential and the Bates static dipole polarization potential \cite{Bates}, $V_{\alpha}(r)$, as in Refs.~\cite{Dolm@Ba+e-,Amusia2015}:
\begin{eqnarray}
U_{\rm C\alpha}(r) =  U_{\rm C}(r) + V_{\alpha}(r),
\label{UCalpha}
\end{eqnarray}
where
\begin{eqnarray}
V_{\alpha}(r)=-\frac{\alpha}{2(r^2 + b^2)^2}.
\label{Valpha}
\end{eqnarray}
Here, $\alpha$ is the static dipole polarizability of C$_{60}$ ($\alpha \approx 850$ $a.u.$ \cite{Amusia_alpha}) and $b$ is generally a free parameter of the order of the fullerene size, $b \approx 8$ $a.u.$ (in Appendix A, we investigate the dependence of electron scattering off
C$_{60}$ and photoionization of $A$@C$_{60}$ on the $b$ parameter and demonstrate that the choice of $b \approx 8$ $a.u.$
is well acceptable).  The
$V_{\alpha}(r)$ potential accounts explicitly for a static dipole polarization potential at large distances from the target  but implicitly for some electron correlation and other multipolar excitations at smaller distances.
The $\psi_{i}({\bm x})$ wavefunction of the $i$'s outgoing electron is then obtained from Eq.~(\ref{eqHF}) where, now, the $U_{\rm C}(r)$ cage potential is replaced by a more complete $U_{\rm C\alpha}(r)$ potential.

\begin{figure}
\includegraphics[width=8cm]{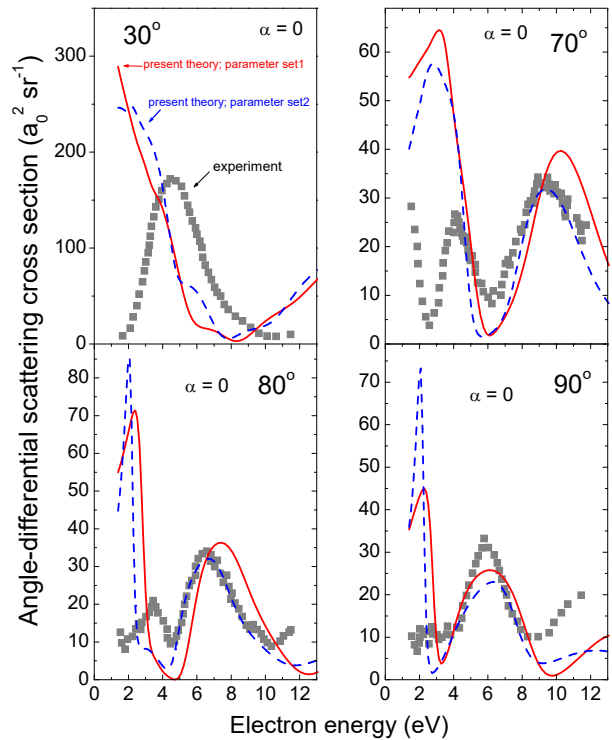}
\caption{The $\rm e + C_{60}$ angle-differential elastic  scattering cross section, $\frac{d \sigma}{d \Omega}$, at given scattering angles $\theta = 30$, $70$, $80$ and $90^{\rm o}$. Solid and dashed lines, results of the present calculation carried out \textit{without} account for polarizability of C$_{60}$ with the use of two sets of the adjustable parameters: $r^{(1)}_{0} \approx 5.8$, $\Delta^{(1)} \approx 1.9$ and $U^{(1)}_{0} = 0.302$ $a.u.$ (referred to
as the ``set$1$'', solid line) and $r^{(2)}_{0} \approx 5.26$, $\Delta^{(2)} \approx 2.91$ and $U^{(2)}_{0} = 0.260$ $a.u.$ (referred to as the ``set$2$'', dashed line); $32$ partial electronic waves with the orbital momentum $\ell$ up to $\ell = 31$ were accounted in the calculation. Solid squares, experiment \cite{Tanaka}.}
\label{DCS0}
\end{figure}

\begin{figure}
\includegraphics[width=8cm]{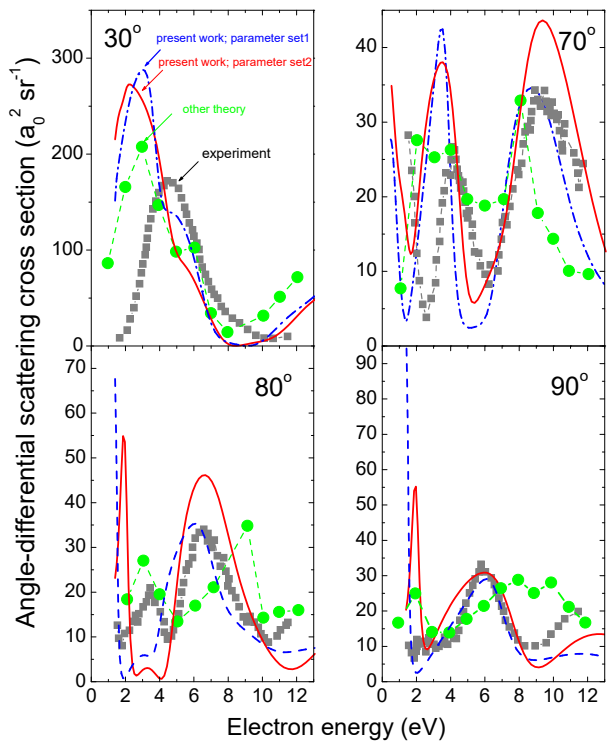}
\caption{The ${\rm e + C_{60}}$ angle-differential elastic  scattering cross section, $\frac{d \sigma}{d \Omega}$, at given scattering angles $\theta = 30$, $70$, $80$ and $90$ degrees. Solid and dashed lines, results of the present
calculation carried out \textit{with} account for polarizability of C$_{60}$ ($\alpha = 850$, $b=8$ $a.u.$) with the use of the set$1$ and set$2$ of the parameters: $r^{(1)}_{0} \approx 5.8$, $\Delta^{(1)} \approx 1.9$ and $U^{(1)}_{0} = 0.302$ $a.u.$ (solid line) and $r^{(2)}_{0} \approx 5.26$, $\Delta^{(2)} \approx 2.91$ and $U^{(2)}_{0} = 0.260$ $a.u.$ (dashed line), respectively; $32$ partial electronic waves with $\ell$ up to $\ell = 31$ were accounted in the calculation. Solid squares, experiment \cite{Tanaka}. Solid circles, results of the non-empirical theory \cite{Gianturco}.}
\label{DCS}
\end{figure}

Let us see, before proceeding to further development of theory in this paper, whether approximating the potential of a polarizable C$_{60}$ by Eqs.~(\ref{UCalpha}) and (\ref{Valpha}) is a viable approximation at all. To reach the goal, let us calculate the angle-differential elastic scattering cross section for ${\rm e +  C_{60}}$ collision, $\frac {d \sigma}{d \Omega}$:
\begin{eqnarray}
\frac{d\sigma}{d\Omega} = \frac{1}{k^{2}}\sum^{\infty}_{\ell,\ell'=0}(2\ell+1)(2\ell'+1)\sin\delta_{\ell}\sin\delta_{\ell'} \nonumber \\
\times \cos(\delta_{\ell} - \delta_{\ell'})P_{\ell}(\cos\theta)P_{\ell'}(\cos\theta).
\label{dsigma}
\end{eqnarray}
Here, $\theta$ and $\Omega$ are the scattering angle and solid angle, respectively, $P_{\ell}(\cos\theta)$ is the Legendre polynomial of the $\ell$th order.
In the calculation, let us use the $U_{\rm C}$ potential, Eq.~(\ref{UCalpha}), and compare the thus obtained calculated data with the available experimental results \cite{Tanaka} and results of a non-empirical theory \cite{Gianturco}. In addition let use two different sets of the adjustable parameters
$r_{0}$, $\Delta$ and $U_{0}$ for the $U_{\rm C\alpha}$ cage potential, Eq.~(\ref{SWP}). One of the two sets of the parameters, the ``set$1$'', is $r^{(1)}_{0} \approx 5.8$, $\Delta^{(1)} \approx 1.9$, $U^{(1)}_{0} = 0.302$ $a.u.$, as in, e.g.,
Refs.~\cite{DolmAQC,Deshmukh@TD} and references therein, the other one, the ``set$2$'', is $r^{(2)}_{0} \approx 5.26$, $\Delta^{(2)} \approx 2.91$, $U^{(2)}_{0} = 0.260$ $a.u.$, as in, e.g., Refs.~\cite{Dolm@Ba+e-,Winstead} and references therein.
Corresponding calculated data are depicted in Figs.~\ref{DCS0} and \ref{DCS}.

A surprising results is that the simple model utilized in the present work provides a reasonable agreement with experiment \cite{Tanaka} [even without account for polarizability of C$_{60}$ ($\alpha =0$), except for angles $\theta = 30$ and $70^{\rm o}$, Fig.~\ref{DCS0}]. The account for C$_{60}$ polarizability, Fig.~\ref{DCS}, corrects the situation considerably, making the agreement between the present theory and experiment be very reasonable; the agreement, the author dares to state,
is, in the whole, even somewhat better than that between the non-empirical theory \cite{Gianturco} and experiment. One would have been too naive to have expected more from the utilized simple modelling of C$_{60}$, and yet its viability is obvious. As for the comparison of results, obtained with the use of the parameter-set$1$ and parameter-set$2$ ($r_{0}$, $U_{0}$ and $\Delta$), it is difficult to give an unambiguous preference to the one over the other; both lead to about the same agreement with experiment, although it looks to the author that the choice of the parameter-set$1$ produces the lowest-energy structures in  $\frac{d \sigma}{d \Omega}$, the traces of which are seen in experiment, whereas the parameter-set$2$ does not. At any rate, the discussed calculated results prove the applicability of the $U_{\rm C\alpha}$ potential, Eq.~(\ref{UCalpha}), to tackling a problem of polarizable C$_{60}$. Note that, obviously, by a fine-tuning of the parameters $r_{0}$, $U_{0}$ and $\Delta$, as well the polarization parameters $\alpha$ and $b$, one can achieve a yet better agreement between theory and experiment than the present one.

\begin{figure}
\includegraphics[width=8cm]{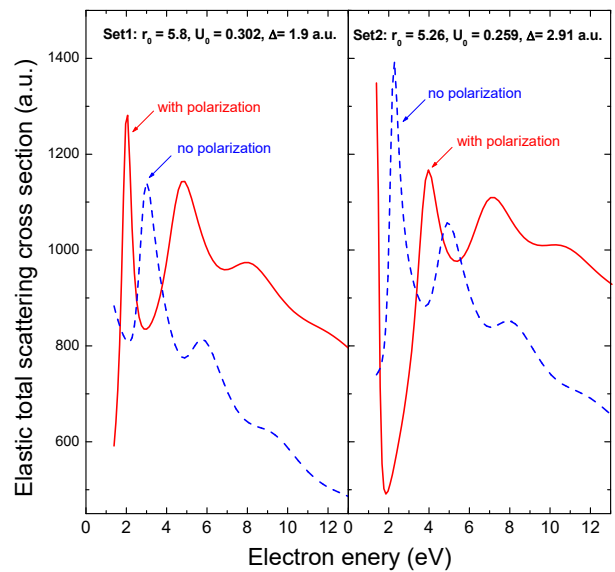}
\caption{ ${\rm e + C_{60}}$ total elastic scattering cross section calculated with and without account for polarizability of C$_{60}$. Solid line (dashed line) - results obtained with the use of the parameter-set$1$ (parameter-set$2$) in the calculation, respectively; $32$ partial electronic waves, with the orbital momentum $\ell$ up to $\ell = 31$, were accounted in the calculation.}
\label{sigmas}
\end{figure}

To let the reader get a better feeling about the difference between a static (``frozen'', $\alpha=0$) and
polarizable ($\alpha \neq 0$) C$_{60}$ fullerene cage, in Fig.~\ref{sigmas} depicted are the $\rm e + C_{60}$ elastic total scattering cross sections calculated with  and without  account for polarizability of C$_{60}$ in the above developed approximations. One can see that the effects of C$_{60}$ polarization alter the scattering cross section considerably; this was previously noted in work \cite{Amusia2015} as well (there, the present author thinks, those
authors mistyped the unit of measurement of the potential depth $U_{0}$ stating it as $U_{0} = 0.52$ $a.u.$ rather than $U_{0} = 0.52$ Ry as it should have been).

Lets us return back to the main topic of the paper -- photoionization of $A$@C$_{60}$ -- and let us build a yet more complete potential of C$_{60}$ felt by the outgoing photoelectron by accounting for the effect termed ``interior static polarization of C$_{60}$'' (``$\zeta$-polarization'')  \cite{DolmPRA2010}.
The quintessence of the
effect is that the ion-remainder, $A^{+}$, once the photoionization
has taken place and the photoelectron is produced,
polarizes the C$_{60}$ cage  much in the
same manner as if it were a conducting shell. This makes the
potential ``felt'' by the outgoing photoelectron be different from $U_{\rm C}$ or $U_{\rm C\alpha}$
both within the wall of the C$_{60}$ cage itself ($r_{0} < r < r_{0}+\Delta$)   and  in the
hollow interior of C$_{60}$, $r < r_{0}$. In a sense, the effect of the interior static
polarization of C$_{60}$ by a vacancy in the endohedral atom
is equivalent to the above discussed relaxation of the atomic core in response
to photoionization of the atom. The interior static polarization of the C$_{60}$ cage by the
ion-remainder $A^{+}$ results in the emergence of an additional potential, $V_{\zeta}(r)$, that acts on a photoelectron,
approximated as \cite{DolmPRA2010}:
\begin{eqnarray}
 V_{\zeta}(r)=\left\{\matrix {
\frac{\zeta}{r_{0}} - \frac{\zeta}{r_{0}+\Delta}, & \mbox{ if $r < r_{0}$} \nonumber \\
 \frac{\zeta}{r} - \frac{\zeta}{r_{0}+\Delta}, &  \mbox{if $r_{0} < r < r_{0}+\Delta$} \nonumber \\
0 & \mbox{otherwise.}
} \right.
\label{Vpzeta}
\end{eqnarray}
Here, $\zeta = 0$ or $1$ if the static polarization is ignored or
accounted, respectively, and the first and second terms in the
upper line of this equation are the constant static potentials
in the inner region due to the negatively charged inner and
positively charged outer surfaces of the C$_{60}$ cage caused by $A^{+}$, respectively.

In order to calculate the $\psi_{i}({\bm x})$ photoelectron wavefunction with account for both interior static polarization of C$_{60}$ by $A^{+}$ and polarization of C$_{60}$ by a photoelectron,
we replace the $U_{\rm C}$ potential in Eq.~(\ref{eqHF}) by a new potential, $U_{\rm C\alpha\zeta}$, which is the sum of all three potentials above:
\begin{eqnarray}
U_{\rm C\alpha\zeta}(r) = U_{\rm C} +V_{\alpha} + V_{\zeta}.
\label{Ualphazeta}
\end{eqnarray}

Alternatively, in order to calculate the $\psi_{i}({\bm x})$ photoelectron wavefunction with account for interior static polarization of C$_{60}$ alone (i.e., ignoring polarization of C$_{60}$ by a photoelectron),
we replace the $U_{\rm C}$ potential in Eq.~(\ref{eqHF}) by the potential $U_{\rm C\zeta}$ which is the sum of only two potentials:
\begin{eqnarray}
U_{\rm C\zeta}(r) = U_{\rm C} + V_{\zeta}.
\label{Uzeta}
\end{eqnarray}

In the present work, we investigate $A$@C$_{60}$ photoionization  with account for each of the above defined approximations separately as well as cumulatively in order to
elucidate their importance in the aim of the present study.

\section{Results and discussion}

\subsection{Photoionization cross sections of $A$@C$_{60}$}

We start from the study of the importance of atomic relaxation in addition to the impact of the $U_{\rm C}$ cage potential on $\rm 4d$ photoionization of Xe@C$_{60}$. This means that we employ GRPAE theory, i.e.,
we first calculate the  ground-state wavefunctions, $P_{n_{j}\ell_{j}}^{\rm +gr}(r)$, of the  Xe$^{+}(...{\rm 4d}^{9}...)@C_{60}$ ion with the help of Eq.~(\ref{eqHF}) where only the $U_{\rm C}$ cage potential is retained. Then, we solve this equation for the wavefunctions of the excited states, $P_{n'_{j} \ell'=\rm f}^{\rm +}(r)$ and $P_{n'_{j} \ell=\rm p}^{\rm +}(r)$, in the self-consistent field of the Xe$^{+}@{\rm C_{60}}$ ion. These excited functions are then plugged into the RPAE equation, Fig.~\ref{fig1}, thereby determining the photoionization amplitudes, cross sections, \textit{etc}, in the GRPAE approximation (with account for atomic relaxation). We also calculate these photoionization quantities in the RPAE approximation, as in Ref.~\cite{DolmCCR}, i.e., with the use of the excited wavefunctions, $P_{n'_{j} \rm f}(r)$ and $P_{n'_{j} \rm p}(r)$, calculated in the field of the frozen atomic core whose wavefunctions are the same as in the neutral atom, Xe@C$_{60}$. The comparison between calculated RPAE and GRPAE results will let us to elucidate the role of atomic-core relation of endohedral Xe, Xe@C$_{60}$. In addition we use two sets of the adjustable parameters
$r_{0}$, $\Delta$ and $U_{0}$ for the $U_{\rm C}$ cage potential, Eq.~(\ref{SWP}). One of the two sets of the parameter is $r^{(1)}_{0} \approx 5.8$, $\Delta^{(1)} \approx 1.9$, $U^{(1)}_{0} = 0.302$ $a.u.$, as in, e.g.,
Refs.~\cite{DolmAQC,Deshmukh@TD} and references therein, the other one is $r^{(2)}_{0} \approx 5.26$, $\Delta^{(2)} \approx 2.91$, $U^{(2)}_{0} = 0.260$ $a.u.$, e.g., Refs.~\cite{Dolm@Ba+e-,Winstead} and references therein.
Now that, finally, reliable experimental data for Xe@C$_{60}$ photoionization \cite{Phaneuf1} are available, we can decisively  determine which of the two sets of parameters suits best for the description of $A$@C$_{60}$ photoionization.

\begin{figure}
\includegraphics[width=8cm]{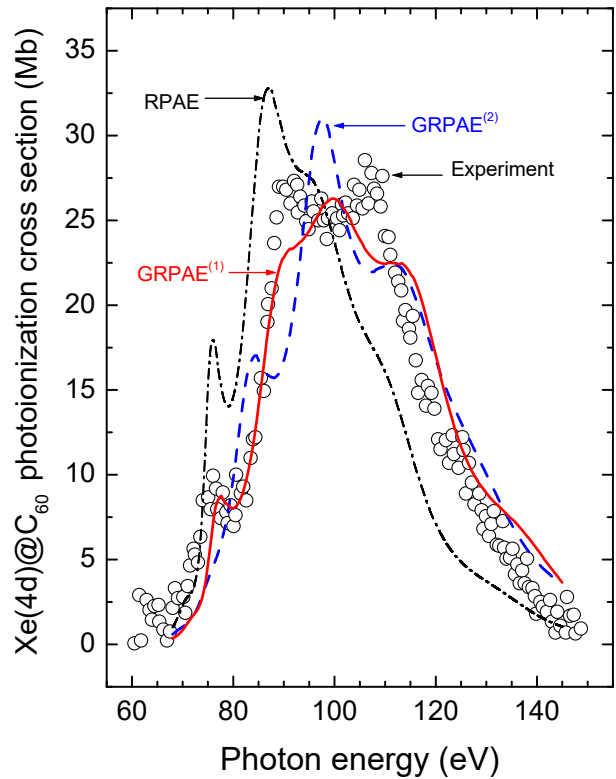}
\caption{The $\sigma_{\rm 4d}$ photoionization cross section of Xe@C$_{60}$ calculated in RPAE (utilizes  $r^{(1)}_{0} \approx 5.8$, $\Delta^{(1)} \approx 1.9$ and $U^{(1)}_{0} = 0.302$ $a.u.$, dashed-dotted line), GRPAE$^{(1)}$ (utilizes  $r^{(1)}_{0}$, $\Delta^{(1)}$ and $U^{(1)}_{0}$, solid line) and GRPAE$^{(2)}$ (utilizes $r^{(2)}_{0} \approx 5.26$, $\Delta^{(2)} \approx 2.91$ and $U^{(2)}_{0} = 0.260$ $a.u.$, dashed line). Neither of these calculated data accounted for any of the polarization effects of C$_{60}$ (see text). Open circles, experiment \cite{Phaneuf1}.}
\label{XeC601}
\end{figure}

Correspondingly calculated RPAE and GRPAE data for the $\rm 4d$ photoionization cross sections of Xe@C$_{60}$ ($\sigma_{\rm 4d}^{\rm RPAE}$ and $\sigma_{\rm 4d}^{\rm GRPAE}$, respectively) are plotted in Fig.~\ref{XeC601}.

The emerging energy-dependent oscillatory structures in $\sigma_{\rm 4d}$, like, e.g., the
 resonant oscillation near $80$ eV, are the confinement resonances \cite{DolmCCR} brought about by the presence of the C$_{60}$ confinement.
 One can conclude from Fig.~\ref{XeC601} that the best agreement with experiment is produced by the GRPAE$^{(1)}$ calculation, i.e., both with account for atomic relaxation of Xe@C$_{60}$ and the use of the set of
$r^{(1)}_{0} \approx 5.8$, $\Delta^{(1)} \approx 1.9$ and $U^{(1)}_{0} = 0.302$ $a.u.$ in the calculation. Hence, the atomic relaxation effect is as important in Xe@C$_{60}$ photoionization as in photoionization of free Xe \cite{Amusia}. As was noted elsewhere above, this fact was not a priory certain, since the C$_{60}$ confinement may make correlation to work differently, even in the opposite way compared to free atom \cite{JPCVKD}. We,
therefore, will discard RPAE calculations from further studies of Xe@C$_{60}$ in the present paper. We also discard the use of the set of $r^{(2)}_{0}$, $\Delta^{(2)}$ and $U^{(2)}_{0}$ from this study, since the GRPAE$^{(1)}$ (to be referred to as GRPAE throughout the rest of the paper) calculation produces better results than the GRPAE$^{(2)}$ calculation.

A fascinating results of the present GRPAE calculation is  that calculated $\sigma_{\rm 4d}^{\rm GRPAE}$ is in a strikingly good agreement with experiment, except for, and yet reasonable agreement, at the very top of  $\sigma_{\rm 4d}^{\rm GRPAE}$; this proves the viability of the simple model utilized in the present study.

\begin{figure}
\includegraphics[width=8cm]{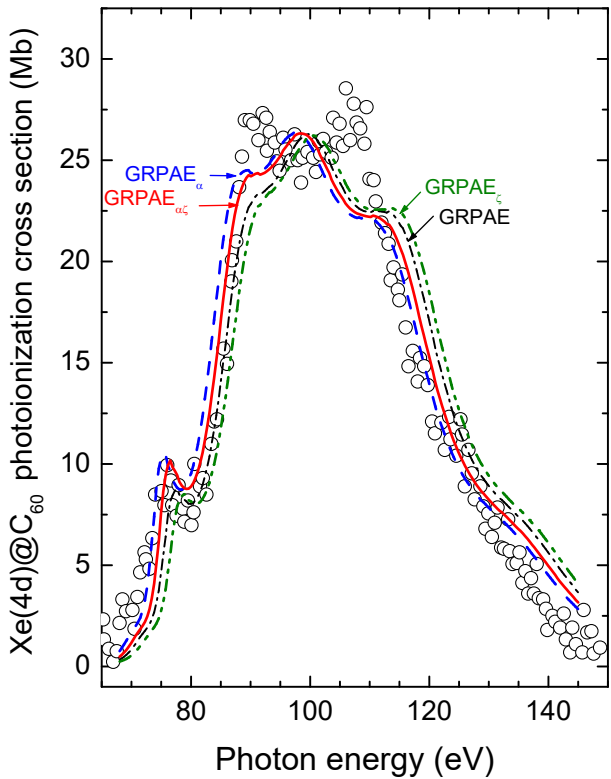}
\caption{The $\sigma_{\rm 4d}$ photoionization cross section of Xe@C$_{60}$ calculated in various approximations. Dash-dot, GRPAE. Dash, GRPAE$_{\alpha}$ (with account for polarization of C$_{60}$ by a photoelectron, only).
Dash-dot-dot, GRPAE$_{\zeta}$ (with account for interior static polarization of C$_{60}$ by Xe$^{+}$, only). Solid, GRPAE$_{\alpha\zeta}$ (with account for both polarization of C$_{60}$ by a photoelectron and interior static polarization of C$_{60}$ by Xe$^{+}$). Open circles, experiment \cite{Phaneuf1}.}
\label{XeC602}
\end{figure}

The above found agreement between theory and experiment brings in a question of whether the account of the impact of polarization of C$_{60}$ by an outgoing photoelectron ($\alpha$-polarization) and/or the effect of interior polarization of C$_{60}$ by Xe$^{+}$
($\zeta$-polarization)
will improve or worsen the GRPAE data. We sought for the answer to this question by carrying out a step by a step study, i.e., by performing GRPAE calculations in three different approximations where the C$_{60}$ potential
was approximated first by the $U_{\rm C\alpha}$, then by $U_{\rm C\zeta}$ and, finally, by $U_{C\alpha\zeta}$ potentials, Eqs.~(\ref{UCalpha}), (\ref{Uzeta}) and (\ref{Ualphazeta}), respectively.
 Corresponding GRPAE calculations will be labelled as GRPAE$_{\alpha}$, GRPAE$_{\zeta}$ and GRPAE$_{\alpha\zeta}$, respectively, and calculated
 photoionization cross sections will be labelled accordingly as well. Calculated results, obtained in the framework of these different approximations, are depicted in Fig.~\ref{XeC602}.

Calculated $\sigma_{\rm 4d}^{\rm GRPAE_{\alpha}}$  is plotted in Fig.~\ref{XeC602} as a dashed line. Surprisingly,
in contradiction to the expectation, based on a large value of the dipole static polarizability of C$_{60}$, the impact of polarization of C$_{60}$ by a photoelectron on the  cross section appears to be relatively weak. Indeed,   $\sigma_{\rm 4d}^{\rm GRPAE_{\alpha}}$ differs from  $\sigma_{\rm 4d}^{\rm GRPAE}$ (dashed-dotted line) primarily by being, as it were, shifted
by mere $2$ eV towards lower photon energies  without practically changing its overall magnitude (a part of the cross section that ``moved'' beyond the $\rm 4d$ threshold is, naturally, cut off the rest of the cross section). In the calculation, we used $b = 8$ for the calculation of the polarization potential, $V_{\alpha}$. In Appendix B, we investigate the dependence of the cross section on various values of the $b$-parameter and show that
 $b = 8$ is a reasonable choice.

Let us focus the attention of the reader on that fact that the agreement between  $\sigma_{\rm 4d}^{\rm GRPAE_{\alpha}}$
and experiment is as good as (maybe a bit better than) the agreement between experiment and $\sigma_{\rm 4d}^{\rm GRPAE}$ calculated without account for the polarization effect.

Calculated  $\sigma_{\rm 4d}^{\rm GRPAE_{\zeta}}$ is plotted in Fig.~\ref{XeC602} as a dashed-dotted-dotted line. One can see that the $\zeta$-polarization effect ``shifts'' (or ``stretches'') the photoionization cross section towards higher photon energies (but the starting point for the cross section remains pinned to the threshold energy), i.e., in the opposite
direction relative to the $\alpha$-polarization effect.  However, the impact of $\zeta$-polarization on the photoionization cross section appears to be about the same as the $\alpha$-polarization effect, i.e., weak.
Yet, compared to  $\sigma_{\rm 4d}^{\rm GRPAE_{\zeta}}$, it is $\sigma_{\rm 4d}^{\rm GRPAE_{\alpha}}$ that is in a somewhat better agreement with experiment.

Calculated resultant $\sigma_{\rm 4d}^{\rm GRPAE_{\alpha\zeta}}$ is plotted in Fig.~\ref{XeC602} as a solid line. It looks like $\sigma_{\rm 4d}^{\rm GRPAE_{\alpha\zeta}}$ is in the best agreement (albeit only slightly better) with experiment compared to calculated  $\sigma_{\rm 4d}^{\rm GRPAE}$, $\sigma_{\rm 4d}^{\rm GRPAE_{\alpha}}$ and $\sigma_{\rm 4d}^{\rm GRPAE_{\zeta}}$. In principle, the entire polarization effect can be safely ignored, since
$\sigma_{\rm 4d}^{\rm GRPAE_{\alpha\zeta}}$ does not differ any significantly from  $\sigma_{\rm 4d}^{\rm GRPAE}$. The atomic-core relaxation effect, however, is clearly important and cannot be discarded in the study of the $\rm 4d$ photoionization cross section of Xe@C$_{60}$.

Let us try to understand why is it that polarization of C$_{60}$ by the outgoing photoelectron has only little impact on Xe@C$_{60}$ photoionization, in particular, and could it be the same for other $A$@C$_{60}$ atoms, in general? In our opinion there are two or several reasons why the $\alpha$-polarization impact on $A$@C$_{60}$ photoionization is not so strong as its impact on electron scattering off empty C$_{60}$, Figs.~\ref{DCS} and \ref{sigmas}.

\begin{figure}
\includegraphics[width=8cm]{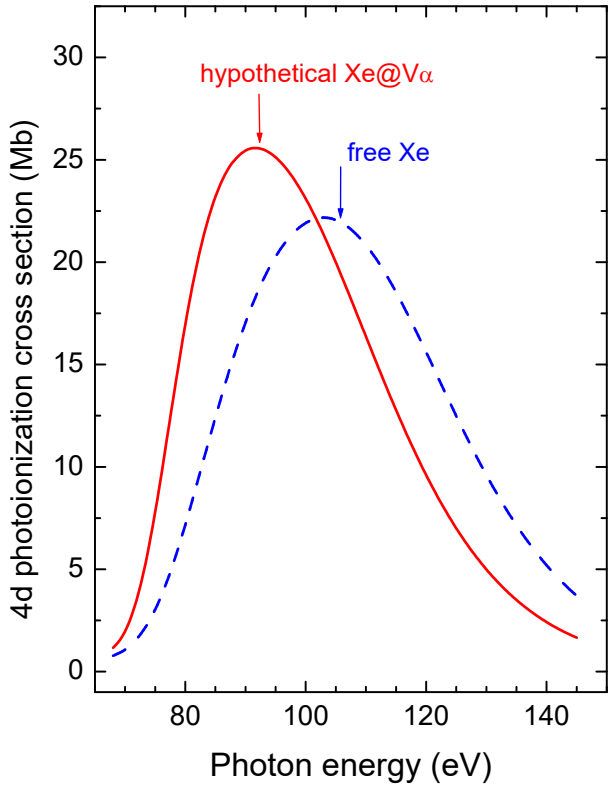}
\caption{The calculated GRPAE $\sigma^{\rm Xe}_{\rm 4d}$ photoionization cross section of free Xe (dashed line) versus the GRPAE $\sigma^{ {\rm Xe}@V_{\alpha} }_{\rm 4d}$ photoionization cross section of a hypothetical Xe@$V_{\alpha}$ atom, i.e., Xe surrounded by the Bates polarization potential, $V_{\alpha}$, Eq.~(\ref{Valpha}), with $\alpha$ being the dipole polarizability of free Xe, $\alpha = 27$ $a.u.$, and $b = r_{\rm 4d} \approx 2.34$ $a.u.$ \cite{Smirnov}.}
\label{hypoXe}
\end{figure}

One of the reasons is associated with the big size of C$_{60}$ itself. The large radius of C$_{60}$  ($\approx 8$ $a.u.$) makes the free \textit{b} parameter in the $V_{\alpha}$ potential, Eq.~(\ref{Valpha}), be large as well,
 $b \approx 8$ $a.u.$. This large value of $b$ makes up for
the large value of $\alpha =850$ $a.u.$, since the denominator of the V$_{\alpha}$ polarization potential, Eq.~(\ref{Valpha}), holds  $b^{4}$, so that  $\frac{\alpha}{b^4}$ is $\approx 0.2$, only. In free Xe, e.g., whose averaged radius
and, thus, $b$ is about $2.34$ $a.u.$ and polarizability $\alpha \approx 27$ $a.u.$ \cite{Smirnov}, the ratio $\frac{\alpha}{b^4} \approx 0.9$ (versus of only $0.2$ for C$_{60}$). Clearly, the large value of $b$ for C$_{60}$ should lessen
 the role of the $\alpha$-polarization potential of C$_{60}$ compared to the ionic potential of $A^{+}$ in the ionized $A$@C$_{60}$, thereby making the role of the former in $A$@C$_{60}$ photoionization relatively insignificant, on the whole. Let us check out this reasoning by considering a hypothetical situation when the Xe atom is surrounded only by the V$_{\alpha}$ static polarization potential, Eq.~(\ref{Valpha}). Correspondingly calculated GRPAE $\rm 4d$ photoionization cross section of such hypothetical Xe is plotted in Fig.~\ref{hypoXe} versus the calculated GRPAE cross section of free Xe.

One can see from Fig.~\ref{hypoXe} than even though polarizability of Xe ($\alpha^{Xe} = 27$ $a.u.$) is much smaller than polarizability of C$_{60}$ ($\alpha = 850$ $a.u.$), its impact on photoionization of the hypothetical Xe@$V_{\alpha}$ atom is noticeably greater than the $\alpha$-polarization impact on photoionization of Xe@C$_{60}$. This is owing to the small $b \approx 2.34$ $a.u.$  cut-off parameter for Xe@$V_{\alpha}$. Indeed, e.g., the maximum of $\sigma^{{\rm Xe}@V_{\alpha}}_{\rm 4d}$ is shifted by about $12$ eV towards lower photon energies compared to the position of the maximum in $\sigma^{\rm Xe}_{\rm 4d}$ of free Xe. This $\alpha$-polarization shift exceeds considerably the mere $2$-eV $\alpha$-polarization shift of   $\sigma_{\rm 4d}^{\rm GRPAE_{\alpha}}$ compared  ``unpolarized'' $\sigma_{\rm 4d}^{\rm GRPAE}$.
In addition, $\sigma^{{\rm Xe}@V_{\alpha}}_{\rm 4d}$ is, overall, noticeably different from
$\sigma^{\rm Xe}_{\rm 4d}$, in contrast to the differences in the cross sections of Xe@C$_{60}$ calculated with and without account for $\alpha$.

\begin{figure}
\includegraphics[width=8cm]{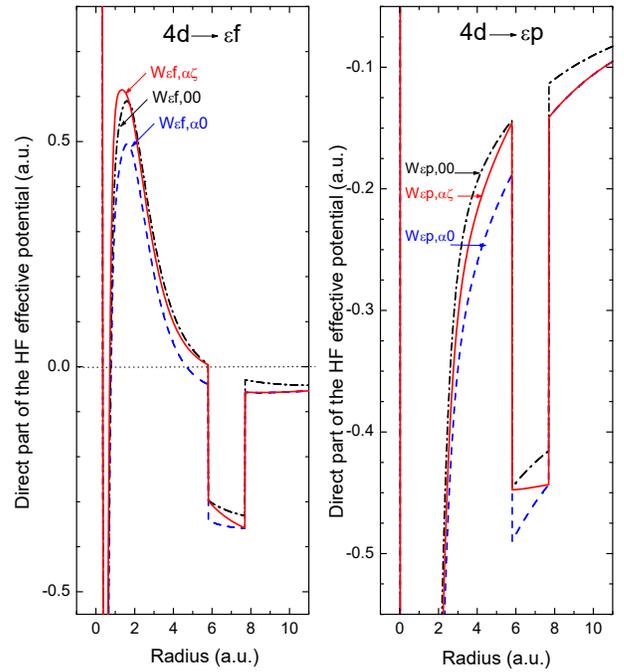}
\caption{Calculated direct parts of the HF potentials felt by the $\epsilon f$ and $\epsilon p$ photoelectrons ejected from Xe@C$_{60}$ upon $\rm 4d$ photoionization, calculated (a) without account for both $\alpha$-polarization and $\zeta$-polarization of C$_{60}$, $W_{\epsilon\ell, 00}$, (b) with account for only $\alpha$-polarization of C$_{60}$, $W_{\epsilon\ell, \alpha 0}$ and, (c),
with account for both $\alpha$-polarization and $\zeta$-polarization of C$_{60}$, $W_{\epsilon\ell, \alpha\zeta}$.}
\label{WdirXe}
\end{figure}

Let us explore how different are at least the direct parts of the HF potentials, felt by a $\epsilon \ell$ photoelectron, due to Xe@C$_{60}$ photoionization, calculated (a) without account for both $\alpha$- and $\zeta$-polarization, $W_{\epsilon\ell, 00}$, (b) with account for $\alpha$-polarization alone, $W_{\epsilon\ell, \alpha 0}$, and, (c)
with account for both $\alpha$-polarization  and $\zeta$-polarization, $W_{\epsilon\ell, \alpha\zeta}$.
Corresponding potentials, calculated for the $\epsilon f$ and $\epsilon p$ photoelectrons due to $\rm 4d$ photoionization
of Xe@C$_{60}$ are plotted in Fig.~\ref{WdirXe}.

One can see form Fig.~\ref{WdirXe} that, indeed the account of only polarization of C$_{60}$ by the outgoing photoelectron does not make the final direct potential, $W_{\epsilon\ell, \alpha 0}$ (dashed line), be noticeably different from
the effective direct potential calculated without account for any polarization of C$_{60}$, $W_{\epsilon\ell, 00}$ (dashed-dotted line). Moreover, when both the dipole $\alpha$-polarization and $\zeta$-polarization are accounted for together, the $W_{\epsilon\ell, \alpha\zeta}$ final potential (solid line) differs only tiny from $W_{\epsilon\ell, 00}$ (dashed-dotted line). This is one of reasons why the polarization impact, especially the combined polarization impact, on the $\rm 4d$ photoionization cross section of Xe@C$_{60}$ results in only a weak alteration of the latter (compared to when the entire polarization impact is ignored).

Another reason for a weak polarization impact on the $\rm 4d$ photoionization cross section of Xe@C$_{60}$ could be as follows. As was shown above, the polarization impact results in only a few-eV, about $2$-eV, ``shift'' of the cross section towards lower photon energies. This results in the part of the cross section, that ``moved'' beyond threshold, being cut off the rest part of the cross section. Hence, if a near-threshold part of the photoionization cross section (a) is large and (b) contains a large part of the total oscillator strength of the continuum spectrum within about $2$ eV starting from threshold,
then the polarization-induced ``shift'' of the cross section to lower energies may affect the cross section noticeably, particularly near threshold. Otherwise, the polarization impact on the cross section should be weak. This
reasoning is fully supported by calculated data for the $\rm 4d$ cross section of Xe@C$_{60}$, obtained with account for $\alpha$-polarization, or $\zeta$-polarization, or both $\alpha$- and $\zeta$-polarization of C$_{60}$.

Let us explore the above-made statement by carrying out calculations for other $A$@C$_{60}$ endohedral atoms but Xe@C$_{60}$. Namely, let us choose the $\rm 2p$ photoionization of Ne@C$_{60}$ (whose $\rm 2p$ photoionization cross section is small at threshold \cite{DolmPRA2010}) and H@C$_{60}$ (whose $\rm 1s$ photoionization cross section is relatively large and changes rapidly near threshold \cite{Balt2,DolmW-S}).

\begin{figure}
\includegraphics[width=8cm]{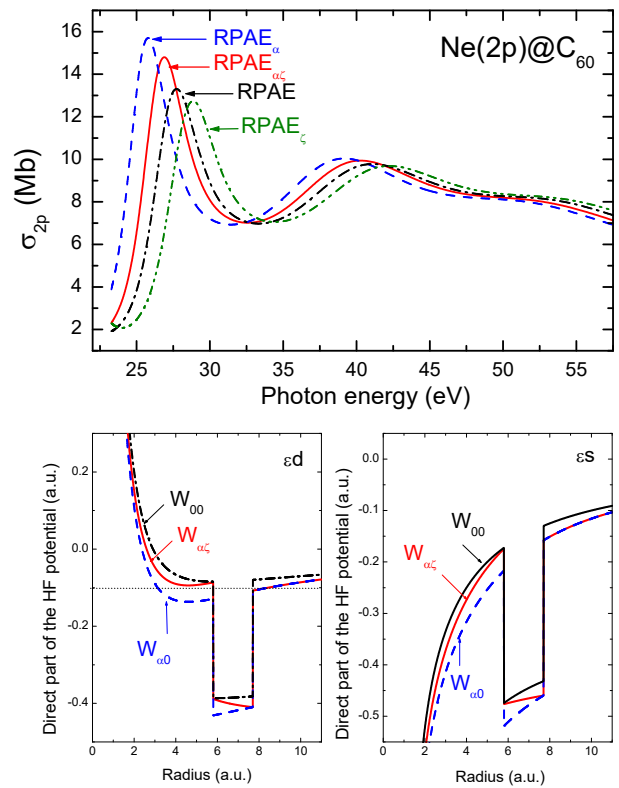}
\caption{Upper panel: the $\sigma_{\rm 2p}$ photoionization cross section of Ne@C$_{60}$ calculated in various approximations (see text). Dashed-dotted line, RPAE (without account for $\alpha$- and $\zeta$-polarization of C$_{60}$:
$\alpha=0$, $\zeta =0$). Dashed line, RPAE$_{\alpha}$ (with account for $\alpha$-polarization of C$_{60}$ alone: $\zeta =0$).
Dashed-dotted-dotted line, RPAE$_{\zeta}$ (with account for $\zeta$-polarization of C$_{60}$ by Ne$^{+}$, only: $\alpha =0$, $\zeta =1$). Solid line, RPAE$_{\alpha\zeta}$ (with account for both $\alpha$- and $\zeta$-polarization of Ne@C$_{60}$: $\alpha=850$ $a.u.$, $\zeta =1$). Lower panel: the $W_{\alpha\zeta}$ direct parts of the HF potential felt by the $\epsilon d$ and $\epsilon s$ photoelectrons ejected upon $\rm 2p$ photoionization of Ne@C$_{60}$:
dash-dot, $W_{00}$ ($\alpha =0$ and $\zeta =0$); dash, $W_{\alpha 0}$ ($\alpha= 850$ $a.u.$, $\zeta=0$); $W_{\alpha\zeta}$ ($\alpha =850$ $a.u.$, $\zeta = 1$).}
\label{NeSIGM}
\end{figure}

Let us start our discussion from the case of Ne@C$_{60}$. Since core-relaxation of Ne upon photoionization is known to be unimportant to the process of photoionization,  we will discard it in our calculations as well, i.e., the calculations will be carried out in RPAE rather than in GRPAE used in the case of Xe@C$_{60}$. Furthermore, like in the case of Xe@C$_{60}$, let us perform a series of calculations: (a) without account for
$\alpha$-polarization of C$_{60}$ (to be labelled as RPAE), (b) with account for $\alpha$-polarization alone (RPAE$_{\alpha}$), (c) with account for $\zeta$-polarization alone (RPAE$_{\zeta}$) and, (d) with account for both
$\alpha$- and $\zeta$-polarization of Ne@C$_{60}$ (RPAE$_{\alpha\zeta}$). Corresponding calculated data for the $\sigma_{\rm 2p}$ photoionization cross section of Ne@C$_{60}$ are depicted in Fig.~\ref{NeSIGM} along with the calculated direct parts of the HF potential felt by the $\epsilon d$ and $\epsilon s$ photoelectrons.

The situation is generally the same as for Xe@C$_{60}$. For $\sigma_{\rm 2p}$, calculated $\sigma_{\rm 2p}^{\rm RPAE\alpha}$ is seen to be ``shifted'' by the effect of $\alpha$-polarization of C$_{60}$  towards threshold without a significant change in its magnitude compared to RPAE $\sigma_{\rm 2p}^{\rm RPAE}$. In contrast, the account of $\zeta$-polarization of C$_{60}$ by Ne$^{+}$ results in
$\sigma_{\rm 2p}^{\rm RPAE\zeta}$ being ``shifted'' (``stretched'') by about $1$ eV to higher photon energies, but, again, without a significant change in its magnitude  compared to
$\sigma_{\rm 2p}^{\rm RPAE}$. When both $\alpha$- and $\zeta$-polarization of C$_{60}$ are taken into account together, they partially cancel out each other. As result,
the resultant $\sigma_{\rm 2p}^{\rm RPAE\alpha\zeta}$ differs little from  $\sigma_{\rm 2p}^{\rm RPAE}$, except for being shifted by approximately $1.5$ eV compared to the main confinement resonance in
$\sigma_{\rm 2p}^{\rm RPAE}$.

Thus, when the photoionization cross section of $A$@C$_{60}$ is small near threshold and does not hold much of  oscillator strength of the continuum spectrum near threshold, the impact of $\alpha$-polarization of C$_{60}$ on the cross section is relatively insignificant, exactly as it has been expected on the basis of the reasoning put forward above.

Let us explore what happens to oscillator strength, $f_{\rm 2p \rightarrow \epsilon}$  (the $\epsilon$ symbol designates the continuum spectrum), concentrated in the continuum spectrum of Ne@C$_{60}$ represented by the $\sigma_{\rm 2p}$ cross section. The needed oscillator strength
is calculated with the help of the well known formula (see, e.g., \cite{Amusia}):
\begin{eqnarray}
f_{n\ell \rightarrow \epsilon } = \frac{c}{2\pi^2}\sum_{\ell' = \ell \pm 1}\int_{I}^{\infty}{\sigma_{n\ell'}(\omega)d\omega}.
\label{OS}
\end{eqnarray}
Here, $c$ is the speed of light, $\omega$ is the photon energy, and $I_{n\ell}$ is the ionization potential of the $n\ell$ subshell of the atom. Calculated data, obtained in the approximations above read:
$f_{\rm 2p \rightarrow \epsilon}^{\rm RPAE} \approx 6.97$, $f_{\rm 2p \rightarrow \epsilon}^{\rm RPAE\alpha} \approx 6.92$ and $f_{\rm 2p \rightarrow \epsilon}^{\rm RPAE\alpha\zeta} \approx 6.95$.
It is evident that account for $\alpha$-polarization of C$_{60}$ ``sends'' a part of oscillator strength from the continuum spectrum into the discrete spectrum of Ne@C$_{60}$. The transmitted part of oscillator
strength is small, being less than $1\%$ of $f_{\rm 2p \rightarrow \epsilon}^{\rm RPAE}$ of the continuum spectrum without account for $\alpha$. This is because the $\sigma_{\rm 2p}^{\rm RPAE}$ cross section
concentrates only a small part of oscillator strength within $2$ eV starting from threshold.

In the two considered examples above, Xe@C$_{60}$ and Ne@C$_{60}$, their cross sections were small and contained only a small part of oscillator strength near threshold compared to the total of oscillator strength of the continuum spectrum.

\begin{figure}
\includegraphics[width=8cm]{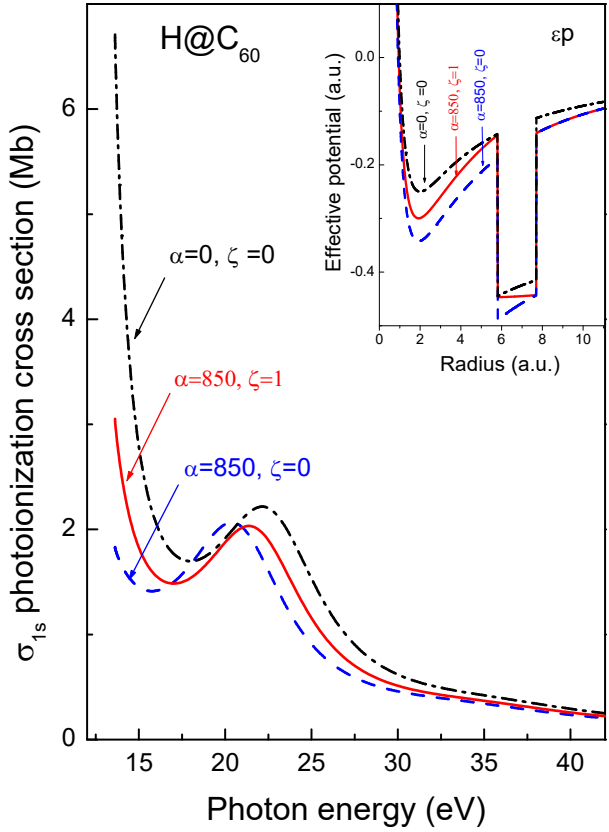}
\caption{Calculated $\rm 1s$ photoionization cross section of H@C$_{60}$, obtained (a) without both account for $\alpha$- and $\zeta$-polarization of C$_{60}$ ($\zeta=0$, $\alpha=0$), $\sigma_{\rm 1s}$, (b)
with account for only $\alpha$-polarization of C$_{60}$, $\sigma_{\rm 1s}^{\alpha}$, and, (b) with account for both $\alpha$- and $\zeta$-polarization of C$_{60}$ ($\zeta = 1$, $\alpha =850$ $a.u.$),
$\sigma_{\rm 1s}^{\alpha\zeta}$,  as marked.
Inset: The effective potential felt by the $\epsilon p$ outgoing photoelectron with the same account for the effects of polarization as for the cross section, as marked.}
\label{figHv2}
\end{figure}

Let us turn to an opposite situation, where, in contrast to Xe@C$_{60}$ and Ne@C$_{60}$, the cross section of an endohedral atom is relatively big near threshold, i.e., a relatively large portion of oscillator strength of the continuum spectrum is concentrated near threshold. With this aim, let us consider $\rm 1s$ photoionization of H@C$_{60}$, since its $\rm 1s$ photoionization cross section, $\sigma_{\rm 1s}$, is known to be relatively large
and changing rapidly near threshold \cite{Balt2,DolmW-S}. Let us calculate $\sigma_{\rm 1s}$ of H@C$_{60}$ with and without account for polarization of C$_{60}$ by a photoelectron.
Also, let us calculate corresponding oscillator strengths, $f_{{\rm 1s} \rightarrow {n\rm p}}$, of the ${{\rm 1s} \rightarrow n \rm p}$ \textit{discrete transitions} from H@C$_{60}$, along with the
total of oscillator strength of the continuum spectrum, $f_{\rm 1s \rightarrow \epsilon p}$. The expectation is that the $\alpha$-polarization effect will, due to the induced ``shift'' of the cross section, effectively cut off the
rapidly changing tail of the cross section near threshold and push a relatively noticeable part of oscillator strength of the continuum spectrum into the discrete spectrum.
Calculated results for $\sigma_{\rm 1s}$ are depicted in Fig.~\ref{figHv2}, along with the $W_{\alpha\zeta}$ potential felt by the outgoing $\epsilon p$ photoelectron.

One can clearly see from Fig.~\ref{figHv2} that the $\alpha$-polarization-induced shift of $\sigma_{\rm 1s}$ effectively cuts off the rapidly-changing  tail of
$\sigma_{\rm 1s}$ near threshold, exactly as was suggested above. Even when interior static polarization is accounted in addition to the $\alpha$-polarization effect, the resultant
near-threshold cross section,  $\sigma_{\rm 1s}^{\alpha\zeta}$ (solid line), still differs noticeably from $\sigma_{\rm 1s}$ calculated without account for any polarization (dash-dot). The percentage differences between the H@C$_{60}$ photoionization cross sections calculated with and without the effects of polarization are stronger than in the case of Xe@C$_{60}$ and Ne@C$_{60}$, exactly as was supposed to happen on the basis of the statements
made above, i.e., the greater near-threshold photoionization cross section and the faster it changes, the greater the polarization impact on the near-threshold cross section.

 Oscillator strengths   of the discrete spectrum, $f_{{\rm 1s} \rightarrow  n\ell}$, and continuum spectrum, $f_{\rm 1s \rightarrow \epsilon p}$, of  H($\rm 1s$)@C$_{60}$  calculated with and without account of polarizability, $\alpha$, of C$_{60}$ are  presented in Table~\ref{Table1}.
\begin{table}
\caption{\label{Table1} Oscillator strengths   of the discrete spectrum, $f_{{\rm 1s} \rightarrow  n\ell}$, and continuum spectrum, $f_{\rm 1s \rightarrow \epsilon p}$, of  H($\rm 1s$)@C$_{60}$  calculated with and without account of polarizability, $\alpha$, of C$_{60}$.}
\begin{ruledtabular}
\begin{tabular}{llr}
{$f_{\rm 1s} \rightarrow n{\rm }p$} & without $\alpha$  & with $\alpha$ \\
%\multicolumn{2}{c}{$\alpha_{4s}$} & \multicolumn{1}{c}{$r_{4s}$} \\
\cline{1-3}

$ \rm 1s \rightarrow 2p$ & 0.146 &  0.205 \\
$\rm 1s \rightarrow 3p$ & 0.289 & 0.427  \\
$\rm 1s \rightarrow 4p$ & 0.124 & 0.027 \\
\cline{1-3}\\
$\sum_{n}f_{{\rm 1s} \rightarrow n{\rm p}}$ & 0.559 & 0.669 \\\\
\cline{1-3}\\
$f_{\rm 1s \rightarrow \epsilon p}$ & 0.352 & 0.257 \\
\end{tabular}
\end{ruledtabular}
\end{table}

 From Table~\ref{Table1}, one can readily conclude that, due to the $\alpha$-polarization effect, the continuum spectrum  loses
$\Delta f = f_{{\rm 1s} \rightarrow \epsilon{\rm p}}^{\alpha=0} -f_{{\rm 1s} \rightarrow \epsilon{\rm p}}^{\alpha \neq 0} \approx 0.1$ of its oscillator strength, i.e., about $30\%$ of its oscillator strength,
whereas the discrete spectrum (approximated by only three discrete transitions in this example) gains about the same amount of oscillator strength, exactly as was suggested above. This loss of oscillator strength by the continuum spectrum of Ne@C$_{60}$ causes a noticeable change to $\sigma_{\rm 1s}$, near threshold, again, as was suggested above.

\begin{figure}
\includegraphics[width=8cm]{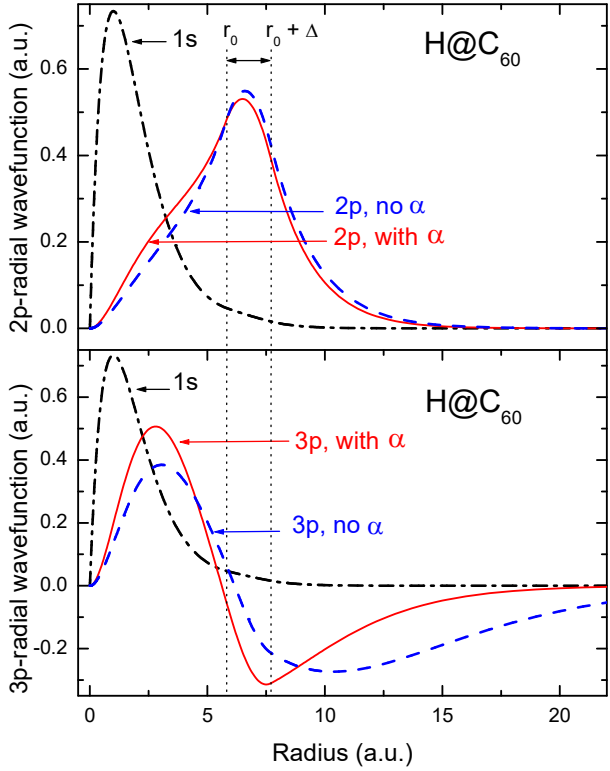}
\caption{Radial wavefunctions of the $\rm 2p$ and $\rm 3p$ excited states of H@C$_{60}$ calculated with and without account of static dipole polarizability, $\alpha$, of C$_{60}$, as well as
the $\rm 1s$ wavefunction,  $P_{\rm 1s}$, of the ground state of H@C$_{60}$, as marked. The vertical dotted lines mark the positions of the inner ($r_{0}$) and outer ($r_{0} +\Delta$) surfaces of the C$_{60}$ shell.}
\label{WFH}
\end{figure}

How could one explain the fact that, for the discrete spectrum,  $f_{{\rm 1s} \rightarrow n{\rm p}}^{\alpha} > f_{{\rm 1s} \rightarrow n{\rm p}}^{\alpha=0}$ (for $n = 2$ and $3$) from a  different point of view?
Since the polarization potential $V_{\alpha}(r)$ is attractive and, particularly, extends into the hollow interior of C$_{60}$, its impact on the excited states must result
in the increased probability density of these states inside the hollow interior of C$_{60}$, i.e., closer to the $\rm 1s$ ground-state orbital of H@C$_{60}$. This, indeed,
is evident from Fig.~\ref{WFH}, where we plotted the calculated radial wavefunctions, $P_{\rm 2p}(r)$ and $P_{\rm 3p}(r)$, of the $\rm 2p$ and $\rm 3p$ excited states of H@C$_{60}$ along with its
ground-state $\rm 1s$ wavefunction, $P_{\rm 1s}$.
Since the overlap between $P_{\rm 1s}$ and  $P_{\rm 2p}$, as well as $P_{\rm 1s}$  and $P_{\rm 3p}$ is greater when the polarization potential is accounted for, then, of course,
$f_{{\rm 1s} \rightarrow n{\rm p}}^{\alpha} > f_{{\rm 1s} \rightarrow n{\rm p}}^{\alpha=0}$.

We now focus the attention of the reader on another interesting fact presented in Table~\ref{Table1}. Namely, in opposition to free H, where
$f_{{\rm 1s} \rightarrow 2{\rm p}} > f_{{\rm 1s} \rightarrow 3{\rm p}}$,  it appears that
$f_{{\rm 1s} \rightarrow 2{\rm p}} < f_{{\rm 1s} \rightarrow 3{\rm p}}$ for H@C$_{60}$, regardless of whether the polarization effect is or is not taken into account. Why?
The answer, again, can be gained from exploring Fig.~\ref{WFH}. One can see that $P_{\rm 2p}$ maximizes \textit{inside the wall} of C$_{60}$,  i.e., in the region of $r_{0} < r < r_{0}+\Delta$, whereas
$P_{\rm 3p}$ wavefunction appears to be significantly drawn into the \textit{hollow interior region} of C$_{60}$,
$r < r_{0}$,  and, as result, maximizes much closer
to the $P_{\rm 1s}$ ground-state  function. Correspondingly, the overlap between $P_{\rm 3p}$ and $P_{\rm 1s}$ is noticeably stronger than the overlap between $P_{\rm 2p}$ and $P_{\rm 1s}$, and,
correspondingly, $f_{{\rm 1s} \rightarrow 3{\rm p}} > f_{{\rm 1s} \rightarrow 2{\rm p}}$, in H@C$_{60}$.
The situation here is reminiscent of, and analogous to, the one discovered for the $\rm 4d$ orbital of Ca@C$_{60}$ and referred to as selective orbital compression \cite{JPCVKD}.

We have shown above, on the example of H@C$_{60}$, that polarization of C$_{60}$ by a photoelectron results in noticeable changes in the photoionization cross section near threshold if the cross section holds
a relatively large portion of oscillator strength and varies rapidly within of about $2$ eV starting from threshold; otherwise changes are small. As was argued above, this is because the $V_{\alpha}$ dipole static polarization
potential of C$_{60}$ is weakened by the large value of the size of the cage [by the large value of the $b$ cut-off parameter in Eq.~(\ref{Valpha})]; this, in turn, weakens the role of the $V_{\alpha}$ polarization potential
compared to the ionic potential of the $A^{+}$ ion-remainder upon $A$@C$_{60}$ photoionization.

\begin{figure}
\includegraphics[width=8cm]{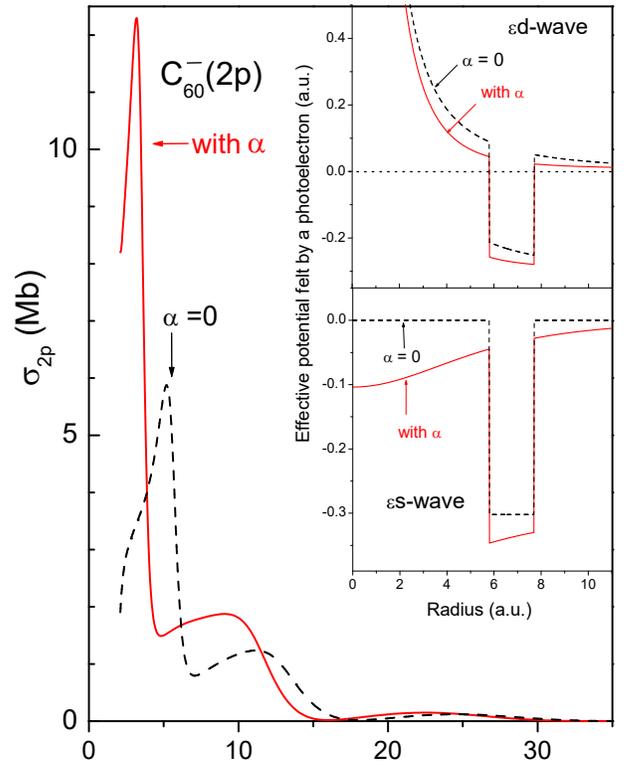}
\caption{The $\sigma_{\rm 2p}$ photodetachment cross sections of a C$_{60}^{-}({\rm 2p})$ hypothetical fullerene anion  calculated with
and without account for static dipole polarizability, $\alpha$, of C$_{60}$,
as marked.}
\label{C602p}
\end{figure}

\begin{figure}
\includegraphics[width=8cm]{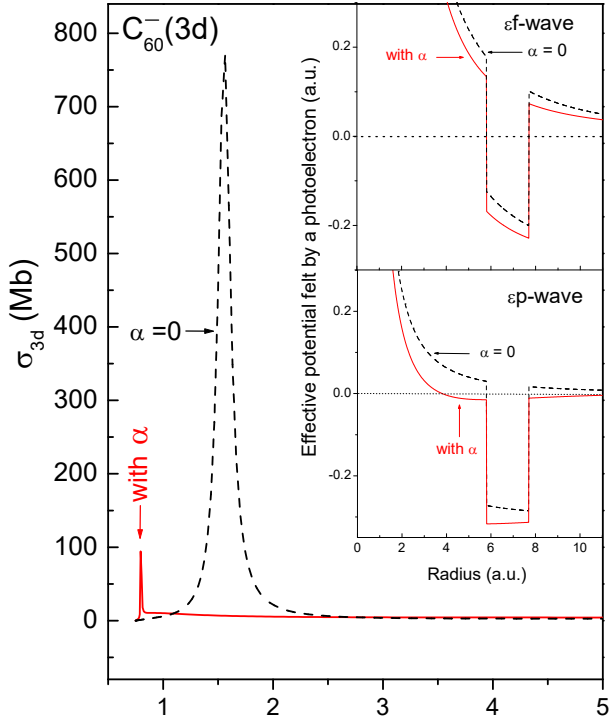}
\caption{The $\sigma_{\rm 3d}$ photodetachment cross sections of a C$_{60}^{-}({\rm 3d})$ hypothetical fullerene anion  calculated with
and without account for static dipole polarizability, $\alpha$, of C$_{60}$,
as marked.}
\label{C603d}
\end{figure}

Let us push the above-made statement to a yet greater extent by considering photoionization of an endohedral system where there is no $A$$^{+}$ ion-remainder in the final state. This, naturally, brings us to the study of photoionization of a fullerene anion, C$_{60}^{-}$. The expectation
is that, in such case, because of the absence of a hampering us strong ionic field in the final state, the effect of $\alpha$-polarization will ``bloom in full swing'', bringing really strong changes to the phodetachment
cross section. To meet the goal, let us consider a C$_{60}^{-}$ fullerene anion in a simplified manner, as in \cite{DolmC60-}, i.e., as a hypothetical fullerene
anion, C$_{60}^{-}({n\ell})$, where the attached electron is bound by the C$_{60}$ cage, $V_{\rm C}(r)$, Eq.~(\ref{SWP}), into a $n\ell$ state which we choose to be either a $\rm 2p$ or $\rm 3d$ state.
Corresponding calculated $\sigma_{\rm 2p}$ and $\sigma_{\rm 3d}$  photodetachment cross sections are plotted in
Figs.~\ref{C602p} and \ref{C603d}, respectively.

One can see that the role of polarization of empty C$_{60}$ by the photodetached  electron on the photodetachment cross section is extremely strong compared to the role of $\alpha$-polarization in $A$@C$_{60}$ photoionization.
It also appears that the polarization effect acts differently in the two considered fullerene anions. In the case of the C$_{60}^{-}(\rm 2p)$ anion, the effect magnifies significantly the photodetachment cross section,
particularly near threshold. It also appears that, as the calculation showed, the oscillator strength of the ``polarized'' continuum spectrum of C$_{60}^{-}(\rm 2p)$ ($f_{\rm 2p \rightarrow \epsilon}^{\alpha} \approx 0.31$) exceeds the
oscillator strength of the continuum spectrum of static  C$_{60}^{-}(\rm 2p)$  ($f_{\rm 2p \rightarrow \epsilon}^{\alpha=0} \approx 0.24$) by about $30\%$. This
$30\%$ gain in oscillator strength of ``polarized'' C$_{60}^{-}(\rm 2p)$ is, apparently, drawn from its discrete spectrum. For C$_{60}^{-}(\rm 3d)$, the situation is opposite. Namely, the
oscillator strength of the  continuum spectrum of static C$_{60}^{-}(\rm 3d)$ ($f_{\rm 3d \rightarrow \epsilon}^{\alpha=0} \approx 1.6$) exceeds considerably oscillator strength of the continuum spectrum of ``polarized''  C$_{60}^{-}(\rm 3d)$  ($f_{\rm 3d \rightarrow \epsilon}^{\alpha} \approx 0.33$). The loss of oscillator strength of the continuum spectrum of C$_{60}^{-}(\rm 3d)$, caused by polarization, results, apparently,
in the transmission of oscillator strength of corresponding continuum spectrum into the discrete spectrum.

All in all, as was expected, due the absence of an ionic potential in the case of C$^{-}_{60}(n\ell)$ photodetachment, the impact of polarization of the C$_{60}$ cage by the outgoing photoelectron on the photodetachment spectrum of a fullerene anion turns out to be incomparably more significant than the same effect in $A$@C$_{60}$ photoionization.

\subsection{Photoelectron angular asymmetry parameter, $\beta_{n\ell}$, and photoionization time delay, $\tau_{n\ell}$}

Let us complete the study of $A$@C$_{60}$ photoionization by exploring the impact of C$_{60}$ polarization, both due to $\alpha$- and $\zeta$-polarization, on the photoelectron angular-asymmetry parameter, $\beta_{n\ell}$, and time delay, $\tau_{n\ell}$, in the $\rm 4d$- and $\rm 2p$-photoionization of
Xe@C$_{60}$ and Ne@C$_{60}$, respectively. Both $\beta_{n\ell}$ and $\tau_{n\ell}$ depend
not only on the absolute values of the photoionization amplitudes, $|D_{n\ell\pm 1}|$, but on the
phases, $\delta_{\ell\pm 1}$, of the matrix elements as well. Furthermore, whereas the total photoionization cross sections, $\sigma_{n\ell}$, are primarily governed by a $n\ell \rightarrow \ell+1$ transition,
the energy dependence of the $\beta_{n\ell}$ and $\tau_{n\ell}$ parameters may be critically affected by a $n\ell \rightarrow \ell -1$ transition itself. Therefore, where the latter transition is affected by the polarization effects strongly, even though it might remain weaker than the dominant $n\ell \rightarrow \ell+1$ transition, there $\beta_{n\ell}$ and $\tau_{n\ell}$ must respond strongly as well, in contrast to the $\sigma_{n\ell}$ total photoionization cross section.

\begin{figure}
\includegraphics[width=8cm]{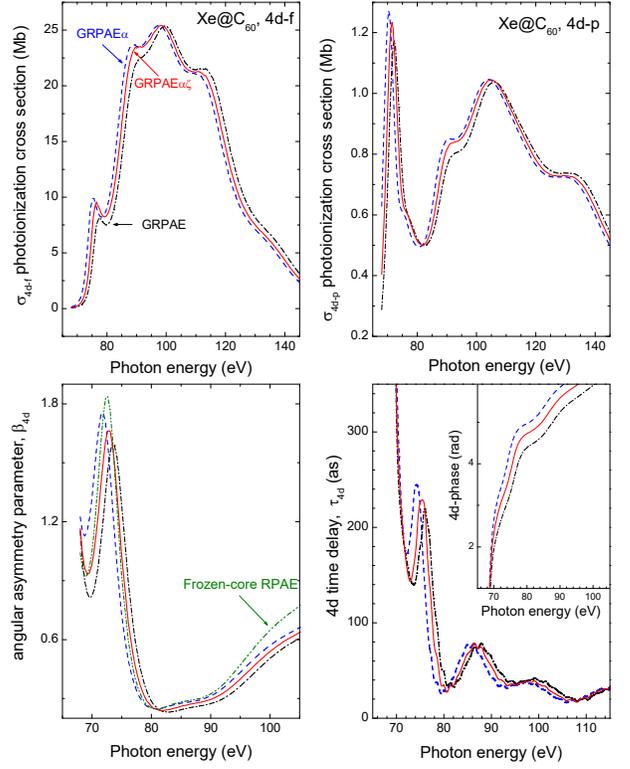}
\caption{The Xe@C$_{60}$ partial $\sigma_{\rm 4d \rightarrow f}$ and $\sigma_{\rm 4d \rightarrow p}$ photoionization cross sections, $\beta_{\rm 4d}$ photoelectron angular-asymmetry parameter, $\tau_{\rm 4d}$  photoionization time delay
and  $\varphi_{\rm 4d}$ phase, Eq.~(\ref{phase}),  of the $\rm 4d$-photoionization amplitude, Eq.~(\ref{Dnl}). Dashed-dotted lines, GRPAE (no account for any polarization of C$_{60}$). Dashed lines, GRPAE$_{\alpha}$ (with account for $\alpha$-polarization, only). Solid lines, GRPAE$_{\alpha\zeta}$ (with account for both $\alpha$- and $\zeta$-polarization of C$_{60}$). Dashed-dotted-dotted line, frozen-core RPAE (without account for core-relaxation of the Xe$^{+}$ ion-remainder and any polarization of C$_{60}$).}
\label{taubetaXe4d}
\end{figure}

\begin{figure}
\includegraphics[width=8cm]{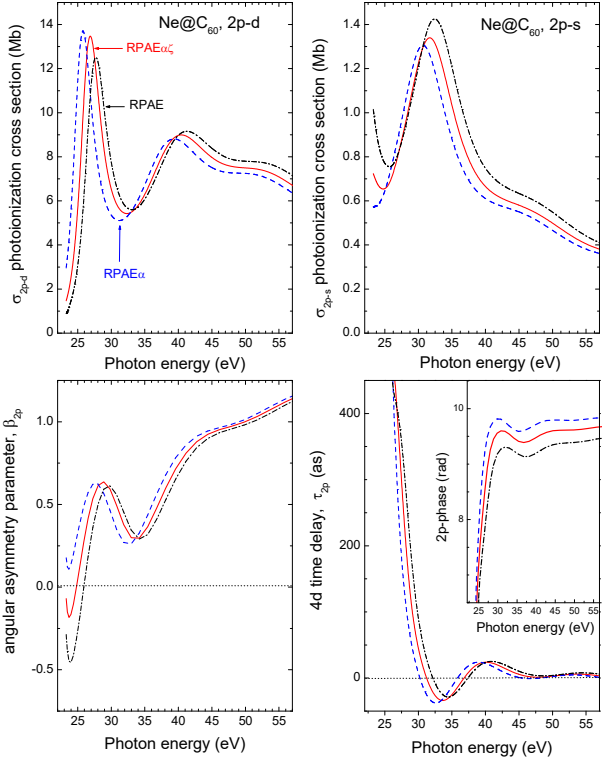}
\caption{The Ne@C$_{60}$ partial $\sigma_{\rm 2p \rightarrow d}$ and $\sigma_{\rm 2p \rightarrow s}$ photoionization cross sections, $\beta_{\rm 2p}$ photoelectron angular-asymmetry parameter, $\tau_{\rm 2p}$ photoionization time delay and $\varphi_{\rm 2p}$ phase, Eq.~(\ref{phase}), of the $\rm 2p$-photoionization amplitude, Eq.~(\ref{Dnl}). Dashed-dotted lines, RPAE (no account for any polarization of C$_{60}$). Dashed lines, RPAE$_{\alpha}$ (with account for $\alpha$-polarization, only). Solid lines, RPAE$_{\alpha\zeta}$ (with account for both $\alpha$- and $\zeta$-polarization of C$_{60}$.}
\label{taubetaNe2p}
\end{figure}

In the present work, the Xe@C$_{60}$'s and Ne@C$_{60}$'s $\sigma_{n\ell \pm 1}$,   $\beta_{n\ell}$, $\tau_{n\ell}$ and the $\varphi_{n\ell}$ phase, Eq.~(\ref{phase}), of the $D_{n\ell}$ photoionization matrix element, Eq.~(\ref{Dnl}), were  calculated in the frameworks of the approximations, referred above to as  GRPAE, GRPAE$_{\alpha}$ and GRPAE$_{\alpha\zeta}$ in application to Xe@C$_{60}$, and  RPAE, RPAE$_{\alpha}$ and RPAE$_{\alpha\zeta}$
in application to Ne@C$_{60}$. Corresponding calculated data are depicted in Figs.~\ref{taubetaXe4d} and \ref{taubetaNe2p}.

One can see that the resultant polarization impact on $\beta_{n\ell}$ and $\tau_{n\ell}$ acts in the same manner as in the total photoionization cross sections: the role of the impact leads to
some change in the phase of the confinement resonance oscillations due to, as it would be, the ``shift'' of the parameters along the energy axis without a strong alteration of their magnitudes.

There are two interesting points to note, though.

First, note, for the case of Xe@C$_{60}$, how all three effects -- the effects of atomic relaxation, $\alpha$-polarization and $\zeta$-polarization of C$_{60}$  -- largely cancel out each other in $\beta_{\rm 4d}$ in the photon
energy region up to about $90$ eV; indeed, the resultant  calculated GRPAE$_{\alpha\zeta}$ $\beta_{\rm 4d}$ differs only tiny (except at the very top of the main maximum) from calculated RPAE $\beta_{\rm 4d}$, obtained without account for either of the effects in question.

Second, note how more significantly the $\alpha$-polarization of C$_{60}$ affects $\beta_{\rm 2p}$ of Ne@C$_{60}$, near threshold, compared to that of Xe@C$_{60}$. Indeed, calculated near-threshold
$\beta_{\rm 2p}^{\rm RPAE_{\alpha}}$ (dashed line)
has the opposite sign compare to $\beta_{\rm 2p}^{\rm RPAE}$ (dashed-dotted line) (calculated without account of any polarization of C$_{60}$). This is because the $\alpha$-polarization effect in Ne@C$_{60}$ alters a smaller
$\sigma_{\rm 2p \rightarrow s}$ partial photoionization cross section not only quantitatively but qualitatively as well, near threshold, by cutting-off a relatively deep near-threshold minimum in $\sigma^{\rm RPAE}_{\rm 2p \rightarrow s}$. However, the combined effect of $\alpha$- and $\zeta$-polarization partially restores the near-threshold minimum
in $\sigma_{\rm 2p \rightarrow s}$, to some extent, thereby lessening the entire polarization impact on $\beta_{\rm 2p}$. As a result, the differences between  $\beta_{\rm 2p}^{\rm RPAE}$ (dashed-dotted line) and
 resultant $\beta_{\rm 2p}^{\rm RPAE_{\alpha\zeta}}$ become no longer qualitative but (not dramatically) quantitative, near threshold.

\section{Conclusion}

In the present work, the impact of polarization of C$_{60}$ by the outgoing photoelectron (the ``$\alpha$-polarization'' impact) on $A$@C$_{60}$ photoionization has been detailed.
It has been found that, contrary to the expectation, the role of polarizability of a highly polarizable C$_{60}$ is surprisingly weak in $A$@C$_{60}$ photoionization. This is linked to the big size of the C$_{60}$ and,
as a result, to the big values of the $b$ cut-off parameter, $b \approx 8$ $a.u.$, in the $V_{\alpha}$ polarization potential, Eq.~(\ref{Valpha}). The viability of the use of $V_{\alpha}$ and the value of $b \approx 8$ $a.u.$
in the aims of the study has been proven in the paper as well. As has been argued in the present paper, the big value of $b$  makes the polarization effect to ``give up'' to the ionic potential of the $A^{+}$ ion-remainder emerging upon photoionization of $A$@C$_{60}$, and there has been presented a plenty number of examples supporting this statement in the main text of the paper and Appendices. The result seems to the author of the paper truly counter-intuitive.
All in all, it has been demonstrated in the present study that the account of $\alpha$-polarization in $A$@C$_{60}$ photoionization can be, in principle, ignored, unless one studies near-threshold photoionization of $A$@C$_{60}$ whose photoionization cross section holds a large, or relatively large amount of oscillator strengths near threshold, within about $2$ eV from threshold; then and there the $\alpha$-polarization impact matters to a greater extent.
Furthermore,
it has been demonstrated that the polarization effect results in the transmission of a part of oscillator strength of the continuum spectrum of $A$@C$_{60}$ into its discrete spectrum.

In addition to the effect of polarization of C$_{60}$ by the outgoing photoelectron, we studied and detailed another, less known effect - interior static polarization of C$_{60}$ by the ion-remainder $A^{+}$
emerging upon photoionization of C$_{60}$ \cite{DolmPRA2010}, termed in the present paper as the ``$\zeta$-polarization'' effect. It has been shown that the $\zeta$-polarization effect is about of the same strength as
the $\alpha$-polarization effect but it counter-acts the former; this makes the entire resultant polarization impact on $A$@C$_{60}$ photoionization yet weaker compared to the competing ionic field of the $A^{+}$-ion-remainder.
We come to understanding that accounting for only the $\alpha$-polarization effect or only $\zeta$-polarization effect in $A$@C$_{60}$ would be erroneous; two effects must be accounted altogether, or both ignored, when studying photoionization of $A$@C$_{60}$.

Lastly, correlation-related atomic-core relaxation of the Xe$^{+}$ ion-remainder upon Xe@C$_{60}$ photoionization and its impact of the photoionization process have been investigated in the present paper as well. This was necessary to do in view of the fact that correlation in $A$@C$_{60}$ can work oppositely the way correlation works in free atoms \cite{JPCVKD}. Anyway, it could not be stated with certainty a priory how atomic-core relaxation
of Xe$^{+}$ in Xe@C$_{60}$ will affect Xe@C$_{60}$ photoionization  compared to the atomic-core relaxation effect in photoionization of free Xe. It has been demonstrated in the present paper that the core-relaxation effect is
as important, and cannot be ignored, in the calculation of the Xe@C$_{60}$ $\rm 4d$ photoionization cross section as in the calculation of that of free Xe. On the other hand, however, it has been found, see Fig.~\ref{taubetaXe4d}, that
the impact of correlation-related core-relaxation of Xe$^{+}$ on the $\beta_{\rm 4d}$ photoelectron angular-asymmetry parameter of Xe@C$_{60}$ is practically cancelled out by the cumulative impact of $\alpha$- and $\zeta$-polarization of the C$_{60}$ cage in a  photon energy region to about $85$ eV away from threshold.

\section{Acknowledgements}
The author acknowledges the research grant from the Research Committee of the University of North Alabama. The undergraduate student Carl Parasility is thanked for the help in performing some of preliminary calculations at the
starting stage of this research \cite{Parasility1,Parasility2}.

\appendix

\section{Electron scattering off C$_{60}$ versus the $b$ parameter in the polarization potential, Eq.~(\ref{Valpha})}

Herein, we investigate the dependence of electron scattering off C$_{60}$ and photoionization of $A$@C$_{60}$ on the $b$ parameter in the $V_{\alpha}$ polarization potential,
 Eq.~(\ref{Valpha}), and conclude that the choice of $b \approx 8$ $a.u.$ is well acceptable.

\begin{figure}
 \includegraphics[width=8cm]{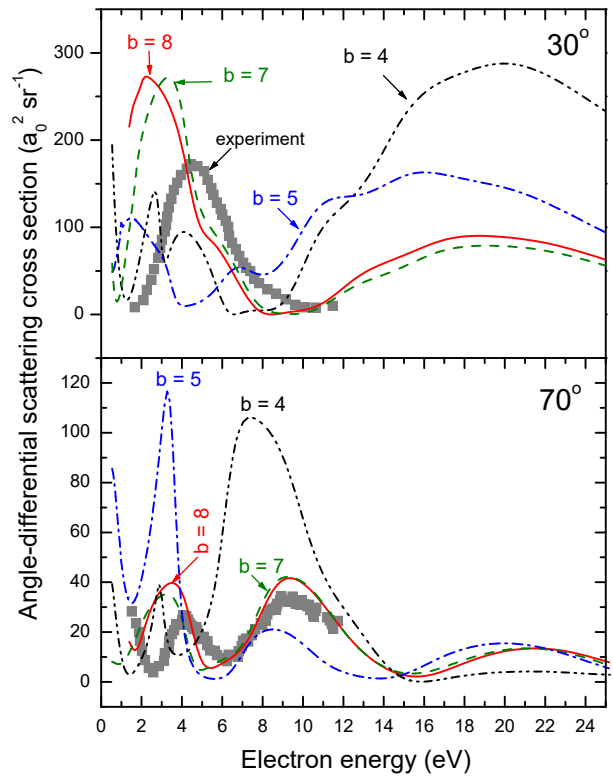}
 \caption{Calculated $\frac{d\sigma}{d\Omega}$ elastic electron differential cross section off C$_{60}$ for scattering angles $\theta = 30$ and $\theta = 70^{\rm o}$, obtained on the basis of  Eqs.~(\ref{UCalpha}) and (\ref{Valpha}), versus the $b$ parameter in Eq.~(\ref{Valpha}). Parameters of the $U_{\rm C}$ square-well potential, Eq.~(\ref{SWP}): $r_{0} = 5.8$, $\Delta = 1.9$ and
 $U_{0} = 0.302$ $a.u.$}
 \label{DCSvsB}
 \end{figure}

 In Fig.~\ref{DCSvsB}, we depict calculated data for the $\frac{d\sigma}{d\Omega}$ elastic electron differential cross section off C$_{60}$ for scattering angles $\theta = 30$ and $\theta = 70^{\rm o}$, obtained on the basis of  Eqs.~(\ref{UCalpha}) and (\ref{Valpha}), versus the $b$ parameter in Eq.~(\ref{Valpha}).
We see that whereas the choice of $b = 7$, $8$ and $9$ leads to about
 the same reasonable agreement with experiment, the decreased values of $b = 5$ and $4$ break the agreement with experiment. We choose $b = 8$ as a reasonable value of $b$ to be used in the calculation of
 elastic electron scattering off C$_{60}$.

\section{Photoionization of Xe@C$_{60}$ versus the $b$ parameter in the polarization potential, Eq.~(\ref{Valpha})}

\begin{figure}
 \includegraphics[width=8cm]{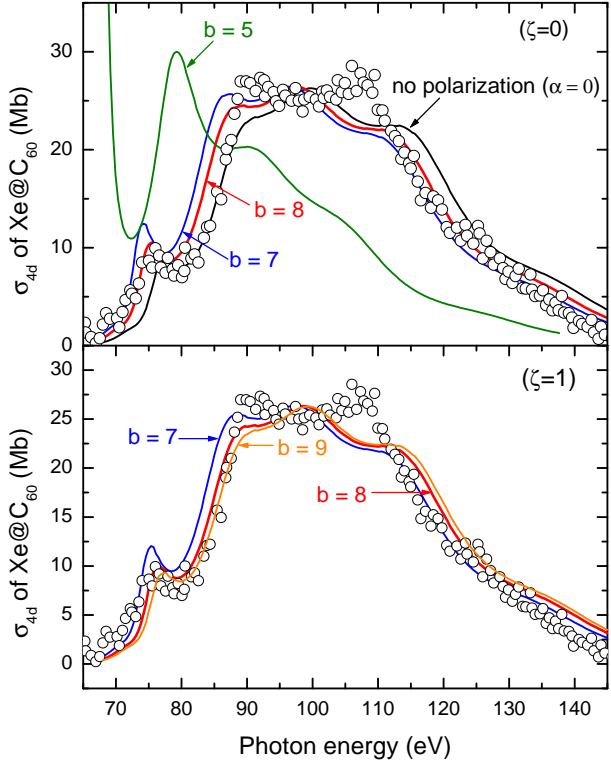}
 \caption{Upper panel: Calculated GRPAE $\sigma^{\rm GRPAE_{\alpha}}_{\rm 4d}$ of Xe@C$_{60}$ with account for the effects of only $\alpha$-polarization of C$_{60}$ (the effect of
 $\zeta$-polarization is omitted, $\zeta =0$), obtained on the basis of
 Eqs.~(\ref{UCalpha}) and (\ref{Valpha}) with the varied $b$ parameter: $b = 8$, $7$ and $5$. Lower panel:  The $\sigma^{\rm GRPAE_{\alpha\zeta}}_{\rm 4d}$ cross section of Xe@C$_{60}$, obtained
 with account for $\alpha$-polarization (with varied $b$:  $b = 9$, $8$, and $7$) in addition to $\zeta$-polarization of C$_{60}$ ($\zeta = 1$). Open circles: experiment \cite{Phaneuf1}. }
 \label{XeC60vsB}
 \end{figure}

 In Fig.~\ref{XeC60vsB}, we depict calculated GRPAE $\sigma_{\rm 4d}$ of Xe@C$_{60}$ with account for the effect of $\alpha$-polarization of C$_{60}$ (i.e., the effect of polarization of C$_{60}$ by the outgoing photoelectron), obtained on the basis of Eqs.~(\ref{UCalpha}) and (\ref{Valpha}) with varied $b$.

 We see that whereas the choice of $b = 8$ and $9$ leads to about the same reasonable agreement with experiment, decreasing values of $b$ below $8$ results in the calculated cross section being ``pushed'' further away from experiment,
 and the choice of a yet smaller $b$, $b =5$, leads to $\sigma_{4d}$ that has nothing in common with experiment. So, we choose $b = 8$, as a reasonable value of $b$ to be used in the
 calculations of $A$@C$_{60}$ photoionization with account for the effect of $\alpha$-polarization of C$_{60}$.

\section{Replacement of the polarization potential, V$_{\alpha}$, by a constant attractive potential inside the hollow interior of C$_{60}$: Close similarity}

\begin{figure}
\includegraphics[width=8cm]{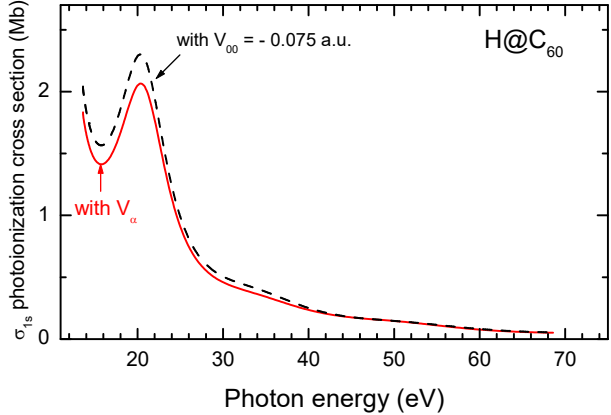}
\caption{The $\sigma_{\rm 1s}$ photoionization cross section of H@C$_{60}$, calculated with the utilization of Eq.~(\ref{UCalpha}), where (a) $V_{\alpha}$ is defined by Eq.~(\ref{Valpha})
with $\alpha =850$ $a.u.$ and $b=8$ (solid line) and (b), dashed line, where $V_{\alpha}$ is replaced by the constant attractive potential inside the hollow interior of C$_{60}$: $V_{\alpha} \rightarrow V_{00} = - 0.075$ $a.u.$, as marked.}
\label{ALPHAvsV00}
\end{figure}

It is interesting to note that the effect of polarization of C$_{60}$ by the outgoing photoelectron, accounted for by the $V_{\alpha}$ polarization potential, Eqs.(\ref{UCalpha}) and (\ref{Valpha}),
can be modelled, to a good approximation, via the replacement of $V_{\alpha}$, in Eqs.(\ref{UCalpha}) and (\ref{Valpha}), by a constant attractive potential, $V_{00}$, inside the hollow interior of C$_{60}$,
$V_{\alpha} \rightarrow V_{00}$:
\begin{eqnarray}
V_{00}(r) = \left\{\matrix {
V_{00} =\mbox{constant} <0, & \mbox{if $0 \le r \le r_{0}$} \nonumber \\
0 & \mbox{otherwise.} } \right.
\label{V00}
\end{eqnarray}
To illustrate this point, our calculated data for the $\sigma_{\rm 1s}$ photoionization cross section
of H@C$_{60}$, obtained with the use of V$_{\alpha}$ and, independently, with the use of $V_{00} = - 0.075$ $a.u.$ in the calculation, are depicted in Fig.~\ref{ALPHAvsV00}.

The observed closed similarity between the results of these two calculations of $\sigma_{\rm 1s}$ of H@C$_{60}$ underlines that fact that, due to the big value of the $b$ cut-off parameter for C$_{60}$,
 the impact of polarization of C$_{60}$ by the outgoing photoelectron (the $\alpha$-polarization effect) on $A$@C$_{60}$ photoionization depends primarily on the inner part of the polarization potential; were the $b$ parameter  small, the situation would have been different. Again, this is illustration of the V$_{\alpha}$ polarization potential, Eq.~(\ref{Valpha}), ``losing'' to the competing ionic potential of the $A^{+}$ ion-remainder in the $A$@C$_{60}$ photoionization process.

\begin{figure}
\includegraphics[width=8cm]{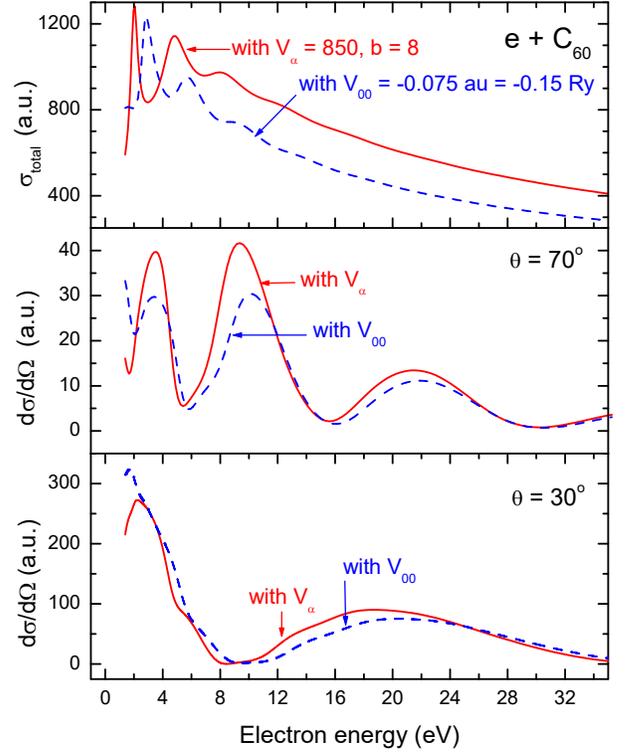}
\caption{The $\rm e + C_{60}$ $\sigma_{\rm tot}$ total and differential, $\frac{d\sigma}{d\Omega}$ (at $\theta = 30$ and $70^{\rm o}$), elastic scattering cross sections
 calculated with the utilization of Eq.~(\ref{UCalpha}), where (a) $V_{\alpha}$ is defined by Eq.~(\ref{Valpha})
with $\alpha =850$ $a.u.$ and $b=8$ (solid line) and (b), dashed line, where $V_{\alpha}$ is replaced by the constant attractive potential inside the hollow interior of C$_{60}$: $V_{\alpha} \rightarrow V_{00} = - 0.075$ $a.u.$, as marked.}
\label{eC60vsV00}
\end{figure}

 The differences between calculated data for ${\rm e + C_{60}}$ elastic scattering, obtained with the utilization of the $V_{\alpha}$ polarization potential in the calculation, on the one hand, and $V_{00}$ constant potential,
 on the other hand, are underlined in Fig.~\ref{eC60vsV00}, where we plotted correspondingly calculated total, $\sigma_{\rm tot}$, and differential, $\frac{d\sigma}{d\Omega}$, cross sections.

One can see from Fig.~\ref{eC60vsV00} that the replacement $V_{\alpha} \rightarrow V_{00}$ does change the total scattering cross section noticeably more significantly than it changes the
H@C$_{60}$ photoionization cross section, Fig.~\ref{ALPHAvsV00}, where there is a competing (to both V$_{\alpha}$ or $V_{00}$) ionic potential of the ion-remainder in the latter.

The replacement, though, does not seem to have changed the $\frac{d\sigma}{d\Omega}$ cross sections a lot; this looks to the author as a curious fact in itself.

\section*{References}

\end{document}